\PassOptionsToPackage{table}{xcolor}
\documentclass[12pt, table]{article}
\usepackage{geometry}
\usepackage{authblk}
\usepackage{hyperref}
\usepackage{appendix}
\usepackage{graphicx, wrapfig}
\usepackage{amssymb}
\usepackage{amsmath}
\usepackage{amsthm}
\usepackage{esvect}
\usepackage{epigraph}
\usepackage{fancyhdr}
\usepackage{xcolor}
\usepackage{caption, subcaption}
\usepackage{geometry}
\usepackage{rotating}
\usepackage[table]{xcolor}
\usepackage{booktabs}
\usepackage{color,soul}

\title{\Large \textbf{Methodology for Modelling Token Economies and Performing Event Impact Analysis with DeTEcT}\\ \large Methodology for modelling token economies using the DeTEcT framework and analysing the significance of event's impacts on the token economy}
\author{R. Sadykhov, Dr.G. Goodell and Prof.P. Treleaven}
\affil{University College London}
\date{June 2026}

\begin{document}

\begin{titlepage}
\maketitle
\thispagestyle{empty}
\begin{abstract}

The objectives of this paper are to provide a methodology for applying the
DeTEcT~\cite{DeTEcT} framework to modelling token economies, to formalise the
configuration of the simulation environment, and to introduce an event analysis framework.
A token economy is an economic system that has a unique mechanism for controlling its
monetary supply, and a medium, in the form of a token or currency, for the valuation of
goods and services, the settlement of transactions, and the storage of value. We show the
key decisions that must be made when modelling an economy with the DeTEcT framework and
showcase some numerical methods that can be used in conjunction with the framework to
perform economic simulations. We also propose a framework for analysing and measuring the
impacts of events on an economy, while also developing a procedure to measure the
significance of these impacts.

Throughout the paper, we use Bitcoin as a case study to demonstrate how to apply the
frameworks and tools we proposed here. We show how a model of Bitcoin token economy can be
set up, and how to measure the impacts of Bitcoin's endogenous policies (i.e., BIPs) on
the wealth distribution of its economic participants.

\end{abstract}
\end{titlepage}

\section{Introduction}
Token economies and financial ecosystems have existed in different forms for as long as humanity was around, with cryptocurrencies representing only one of the latest evolution of these. However, regardless of how long and in what form these token economies exist, they always present similar challenges: ``should they be regulated?'', ``who should regulate them?'', ``how do we ensure fair conditions for the participants of these economies?'' or ``how do we protect consumers from being exploited by malicious agents in the economies?''. These are only some of the questions that we need to answer to bring stability, justice and security to token economies for their adoption by the general public.\par
Cryptocurrencies are a good current example of where regulators need to address problems \cite{FCAandBoETokenisation}, as the original idea behind them was not to create a financial instrument with a new risk profile, but rather to create self-sustaining economies, with potential for further ``layered'' financial solutions built on top of them, which only reinforces the narrative that cryptocurrencies are solutions for facilitating regulatory arbitrage.\par
The existence of a cross-border transaction system that facilitates alternative financial solutions with unmeasured risk forces the regulators world-wide to find ways of regulating, which necessitates modelling and understanding these ``token economies'' (as defined by Sadykhov et al \cite{DeTEcT}). For this purpose we have introduced DeTEcT \cite{DeTEcT} as an agent-based framework for modelling the dynamics of wealth distribution in token economies by modelling the interactions between different agents (or groups of agents called \emph{Agent Categories}). Given the topology of an economy and the data available, one can employ DeTEcT to construct a model of the token economy, which we can use to examine how different interactions in the economy impact the movement of wealth in it. This is potentially useful for regulators seeking to model a possible impact that policies, or other fundamental events, might have on the distribution of tokens between the participants of the economy or its compartments.\par
DeTEcT at its core is flexible and can be applied to creating different theoretical models of token economies \cite{DeTEcTExtension}, so in this paper we demonstrate how DeTEcT can be configured for modelling a token economy and analysing the policy introductions in it, while also providing an example of Bitcoin token economy as a case study. We also validate the results that we obtain against an empirical analysis that we have previously conducted \cite{BTCImpactsOfPolicies} to show that the analysis framework proposed here is robust and practical.\par
By modelling a token economy and examining how the dynamics of wealth distribution changes before the policy introduction and after (i.e., before a specific endogenous policy is proposed and after, as we assume that policy changes impact the economy at the time of release rather than at the time of implementation) will expand the regulators' toolkit for controlling wealth dynamics in token economies by suggesting policies that can be implemented, provide an analysis framework that can be used to measure the impact of these policies, and give us a way to model future policy implementations. The objective of this paper is to propose a refinable and repeatable methodology, and to illustrate its application using Bitcoin as a case study.\par

\section{Scope}
\label{Section:Scope}
For us to perform the analysis described above, we first consider different ways that DeTEcT, being a theoretical framework, can be implemented in code. This includes not only the choice of the programming language, but also a choice of numerical methods.\par
Internally, DeTEcT uses a dynamical system with every equation representing the flow of tokens of a specific \emph{Agent Category}, which is a collection of economic participants (i.e., agents) that satisfy some predefined qualifying criteria (e.g., Consumers, Producers, etc). Each of these agent categories can interact with or rotate into other agent categories. Interactions are the movements of wealth between two agent categories in a form of transactions (e.g., a consumer has purchased a good from a producer), whereas rotations are defined as the movement of an agent from one category to another, and therefore, taking his wealth with him to a different category (e.g., a consumer who has previously purchased a good from a producer, now has produced his own good, therefore becoming a producer, and sold it to a different consumer). Given the set of rotation and interaction rates (i.e., the coefficients in the dynamical system that represent the rate of wealth diffusion between two agent categories), it is possible to simulate the evolution of a dynamical system until a certain time horizon to demonstrate how the system behaves under this specific parametrisation within the set time period.\par
Conversely, we can take a certain final state of the wealth distribution between agent categories, and use numerical methods to solve the dynamical system subject to the boundary conditions (i.e., starting and final wealth distribution states that the system must attain to) - we refer to this as \emph{Backward Propagation}. This way we get a set of interaction and rotation rates that can be set in the system to achieve the desired dynamic. Given wealth distribution time slices of a token economy, we can study the parameters of the system and analyse the difference in old and new rates of diffusion between agent categories.\par
For our implementation we use the Rust programming language, because it is a fast, memory-safe, and reliable systems-design language. For the purpose of this paper, we have chosen three numerical methods to test with the backward propagation. They are \emph{Gradient Descent}, \emph{Nelder-Mead}, and our own extension of Nelder-Mead, which adds random jumps if the solver gets stuck in a local minimum. We chose these numerical methods to adapt because:
\begin{itemize}
	\item \textbf{Nelder-Mead} is a very fast optimisation method that does not require a Jacobian or Hessian of the objective function, making it a convenient choice for optimising elaborate functions.
	\item \textbf{Gradient Descent} is the go-to choice for training neural networks so we want to see its performance when running DeTEcT simulations.
	\item \textbf{Nelder-Mead with Random Jumps} is designed to solve a possible issue with Nelder-Mead, where the numerical engine gets stuck on a sub-optimal solution of the dynamical system.
\end{itemize}
The loss functions that we use are \emph{Mean Absolute Error} (i.e., MAE) and \emph{Mean Log Error} (i.e., MLE). Given the choice of numerical engines and loss functions, we then define a method for selecting an appropriate combination of these configurations for further simulations of a specific token economy, as we believe that different numerical techniques may be better suited for different token economy models.\par
Having formalised the configuration of the economy model and simulation environment, we aim to formalise the analysis of impacts of policies, and generally speaking, events, on a token economy, using DeTEcT as the backbone of the test. We design a analysis framework that uses DeTEcT simulations at its core, providing a quantitative measure for the significance of changes attributed to the event, while also delivering tools for structuring and providing additional perspective for further qualitative analysis.\par
Alongside formalising the methodology, we offer a case study to display how this methodology can be applied  to a real token economy. For this purpose we use Bitcoin as it is a well established token economy with an abundance of historical data. Specifically, we use data describing the wealth distribution of Bitcoin (i.e., 10 discrete Gini-like buckets of aggregate wealth of addresses) from January 2009 to August 2025, provided by CoinDesk \cite{CoinDesk}. This implies that the conclusions of our case study tests will be presented with respect to these wealth distribution buckets (i.e., in our examples we will see how economic policy changes impacts the structure of Lorenz curve of the Bitcoin economy). However, this does not imply that the methodologies described here are only applicable to simulate wealth distribution dynamics between Gini-like buckets. If there is a different partition of wealth distribution data in an economy (e.g., imagine a case with a protocol, consumers and service providers being $3$ agent categories, with a wealth distribution data available for these) then we claim that the methodologies presented in this paper are still applicable to model the token economy, as there is no part in this methodology that is specific to Gini-like buckets, but it means that the conclusions drawn are relevant to the economy model defined via that specific data partition. This may be a useful procedure when trying to examine economy from different perspectives, however, we won't expand on it in this paper. Another reason for picking Bitcoin as a case study is that it has a ready available set of endogenous policies called BIPs \cite{BIPs} that we use in our analysis as the economic policies introduced to the token economy.\par
To validate the conclusions we get in our Bitcoin case study, we refer to the empirical analysis we have previously conducted \cite{BTCImpactsOfPolicies}. We use it as a guide line for setting the $6$-month event impact window, which means that at most $6$ months passes from the introduction of a BIP until a statistically significant effect on the Bitcoin wealth distribution. Additionally, we use the set of \emph{Major Economy-Related BIPs} \cite{BTCImpactsOfPolicies} in this paper (i.e., BIPs 32, 42, 50, 141, 341) as the examples of policies whose impact we will analyse using our proposed test framework, and we validate the tests' results aganst the patterns we saw in that empirical analysis. On individual BIPs, we use the \emph{Created} field as the date around which we are testing, as it is the date the specific BIP was created on in the BIPs repository and became officially publicly visible. Note that it does not reflect the date when (and if) the proposition was integrated into a Bitcoin protocol. Therefore, we should see these as policy announcements, as oppose to actions, but financial markets and economies, more often than not, ``move'' on announcements rather than actions themselves. We used this assumption in the empirical analysis \cite{BTCImpactsOfPolicies}, where we used \emph{Created} dates to define the time series of policy changes for the Granger-causality test, and since we use the results of our causal analysis here to test DeTEcT against, we will use this assumption in this paper for consistency of results. Additionally, all BIPs that we examine in this paper have been implemented (i.e., none of them were refused).\par
Furthermore, in that analysis, we use the same Bitcoin wealth distribution dataset as stated above, which does not contain \emph{Control Mechanism} as a distinct wealth distribution bucket as it is a concept that is restricted to DeTEcT framework, so there is no empirical analysis done on this agent category specifically. However, for other buckets we examined the links between \emph{Major Economy-Related BIPs} and wealth distribution dynamics, which revealed that these major economic policy changes tend to impact poorer and mid-tier wealth distribution buckets (yet not the poorest bucket) as there was a statistically significant wealth distribution changes in these buckets following the policy changes; the wealthy buckets did not show any significant statistical relationship with the introduction of policies. This phenomenon potentially can be explained by wealthier Bitcoin addresses having a better awareness of the Bitcoin economy, access to better quality resources and data, and of course the voting power to push through some of this policies, which in turn means that the wealthier addresses have ``priced-in'' the new policy before it was publicly announced.\par
Lastly, for the brevity of the paper, we will not provide DeTEcT definitions as they can be found in \emph{Decentralized Token Economy Theory (DeTEcT)} \cite{DeTEcT} and \emph{DeTEcT: Dynamic and Probabilistic Parameters Extension} \cite{DeTEcTExtension}.

\section{Configuration of Economy Model and Simulation Environment}
\label{Section:TechnicalConfiguration}
In this section we are looking to formalise the parametrisation of DeTEcT simulation engine as well as selection of a numerical optimisation method for finding solutions to backward propagation of a token economy model. Once we described how to apply DeTEcT to create a model of an economy, and defined the process of numerical optimisation method selection, we proceed to apply this methodology to a model of the Bitcoin token economy, where we analyse selection of an appropriate combination of numerical techniques.\par
\subsection{General Methodology: Modelling Token Economy}
\label{Subsection:ModellingTokenEconomy}
In order to provide a well-structured formal methodology, we decided to split it into steps for convenience. Each step describes a separate task, or a number of tasks that are thematically grouped, which need to be completed to apply DeTEcT to modelling real-world economies. Note that at the end of this section we summarise each step in a simplified format.\par
\textbf{[Step 1]} The first stage of modelling using DeTEcT comes in the form of identifying how to partition economy into agent categories. This partitioning we can refer to as agent category taxonomy and it is an exhaustive mapping of all individual agents (i.e., participants) in the economy into thematic buckets with their aggregate wealth assigned to these buckets. The choice of partitioning is important as it defines the perspective that we use to model the economy - the same economy can be described from perspective of rich-and-poor agents, consumers-and-producers, or other, more complex scenarios. Depending on the agent category taxonomy, the simulations and tests  will uncover different economic phenomena despite individual agents retaining their original individuality.\par
\textbf{[Step 2]} Having identified the desired agent categories, it is then necessary to partition wealth distribution data at each available timestamp into these agent categories accordingly. Notice that by having a dataset where each individual agent and his respective wealth are kept as records, as opposed to keeping the aggregate data on agent categories, we can partition the agents into categories in many different ways by reusing the granular agent-wealth data.\par
\textbf{[Step 3]} Forward propagation requires an initial wealth distribution state as a boundary condition, while backward propagation requires boundary conditions to be set in the form of initial and final wealth distribution between different agent categories. When testing for an optimal backward propagation configuration, we run the simulations between the two points in the wealth distribution with the largest difference between them. Usually this turns out to be the oldest data point and the most recent one. The motivation for this is to run the simulations between the most ``different'' data points to put the largest possible stress on the numerical engine in order to understand how well different numerical optimisation methods perform under the most extreme conditions. Note that for most models of real-world token economies we will see relatively small redistribution of wealth between data points that are less than a year apart, which means that most numerical optimisation methods will do the job adequately in those circumstances. However, the objective here is to identify the optimal configuration to run further simulations under any conditions.\par
\textbf{[Step 4]} Now that we have selected an  agent category taxonomy and prepared the data to initiate the simulations, we need to establish the topology of possible interactions and rotations in the economy. By default, we can assume that the economy does not allow any agent categories to interact with each other or rotate into each other (i.e., we set all interaction and rotation rates to $0$), and then we can start adding the \emph{Interaction} and \emph{Rotation Types} that we allow in the economy. Alternatively, we can assume that all \emph{Interaction} and \emph{Rotation Types} are allowed, and proceed to remove the interactions and rotations that we want to restrict. The choice of approach is dependent on the economy being modelled and the associated agent category taxonomy (e.g., in the case of Bitcoin and Gini-buckets, we assume that every individual agent can interact with each other as there is nothing to indicate the opposite, while due to the taxonomy of agent categories we see that the rotations are poorly defined so we set all of them to $0$). Constructing an exhaustive topology is tedious, however, having a good interactions and rotations topology is useful as it further constraints the model and reduces the degrees of freedom making simulations potentially faster and more accurate. Additionally, some economic policies may change the topology, which is a structural difference that we must manually apply to the simulation once it is in effect (e.g., an example of such policy is an embargo of interactions between two specified agent categories - this restrictions sets the interaction rates between these agent categories to $0$, preventing them from ``exchanging'' wealth during the simulation).\par
\textbf{[Step 5]} The steps outlined above are needed for us to construct the mathematical representation of the economy. Recall from Sadykhov et al \cite{DeTEcT} that the generalised form of the dynamical system for $n$ agent categories is:
\begin{equation}
\label{Equation:TechnicalConfiguration:GeneralSystem}
	\frac{\Delta}{\Delta t}[\vv{F}(t)] = \frac{1}{M}\vv{F}(t)\odot[\mathcal{B}\cdot\vv{F}(t)] + \Gamma\cdot\vv{F}(t),
\end{equation}
with $\mathcal{B}$ being the antisymmetric matrix of interaction rates, and $\Gamma$ the matrix of rotation rates where each column sums up to zero. The vector of wealth functions, $\vv{F}(t)$, is the distribution of wealth between the agent categories at time step $t$. For token economies we would add \emph{Control Mechanism} as an extra agent category to represent the tokens held by the protocol, and to some economies we may want to add other specialised agent categories to tailor the model to its real-world counterpart (e.g., we would introduce \emph{Token Dump} if we were modelling Ethereum token economy to represent the zero address used by Ethereum smart contracts to burn token supply). The importance of \emph{Control Mechanism} is that it is assumed to ``own'' all unissued tokens (i.e., it represents the difference between the maximum supply and issued supply at every point in time).\par
At this stage we could add different extensions to our model (e.g., dynamic money supply \cite{DeTEcTExtension}), but for simplicity we only present the standard DeTEcT dynamical system. That being said, if we are looking to configure the simulation environment to run backward propagation simulations, we must ensure that the parameter extension we use must be compatible with backward propagation simulations (e.g., dynamic probabilistic parameter extension cannot be used).\par
\textbf{[Steps 6-8]} Having defined a model for an economy, we can use Euler method to run the forward propagation simulations, which starts at the initial wealth distribution between the agent categories, and forecasts the dynamics of wealth distribution for each agent category over a set time period. This would conclude the configuration of the simulation environment for forward propagation simulations. However, since we are interested in running backward propagation simulations, as we need these in our event analysis framework in Section \ref{Section:EventAnalysis}, we need to do additional work, specifically to select and configure the numerical optimisation methods that we would use to find parameters of our dynamical system in backward propagation.\par
Backward propagation works by running a forward propagation with an initial guess of interaction rates, with the final wealth distribution state compared against the desired one at the final timestamp using a loss function. This means that the system will be simulated from the initial state of wealth distribution using some guess of interaction and rotation rate parameters and Euler method to construct the dynamics of the system, the loss will then be computed by iterating until either the loss is below the required level, or the numerical optimisation method has reached some hard stop that makes it exit the simulation. We equipped all of the numerical optimisation methods that we adapted with two hard stops: maximum number of function calls, and the maximum number of iterations. For the purpose of this study, we have not limited the number of maximum function calls in our case studies, as gradient descent is very ``function call''-hungry even when its number of iterations is relatively modest, but in general one can experiment with the maximum number of function calls for these optimisation methods as well as the maximum number of iterations.\par
As mentioned in Section \ref{Section:Scope}, we adapted three numerical methods for finding a local minimum:
\begin{itemize}
	\item \textbf{Nelder-Mead} \cite{NelderMead}: Numerical method used to find a local optimum (i.e., maximum or minimum) in a multi-dimension space without requiring any derivatives (i.e.,  Jacobian or Hessian) of the objective function. This method may converge on a non-stationary point of the system.
	\item \textbf{Nelder-Mead with Random Jumps}: Nelder-Mead method with an extension that on every iteration checks if the difference of current and previous loss is within a certain interval, and if it is, then the number of ``similar'' losses in a sequence is incremented by $1$. If the number of ``similar'' losses reaches a pre-defined maximum number, a new random guess for the parameters is generated and the optimisation continues from this point, while the number of ``similar'' losses resets to $0$. This approach helps us to soft-reset the numerical optimisation if it gets stuck in a tight range of losses, without exiting the optimisation process.
	\item \textbf{Gradient Descent} \cite{GradientDescent}: Numerical method used to find a local minimum by moving along the space in the opposite direction to the gradient at the current point, and therefore, moving in the direction of a local minimum.
\end{itemize}
Nelder-Mead is one of the most popular numerical optimisation methods as it doesn't rely-on a Jacobian or Hessian of the objective function to be provided as part of the optimisation problem. It is also very fast, which is why we chose to adapt it. We were worried that it might get stuck in a local minima, so we introduced a variant of Nelder-Mead method which incorporates random jumps to reposition and reset its simplex for further iterations. The inclusion of gradient descent here is due to the fact that it is a commonly used minimisation technique too, which gained popularity because of its application to neural networks, and we are curious with regard to how well it will perform in the context of DeTEcT simulations. Note that we did not add more numerical optimisation methods as we had to adapt these to work in Rust in conjunction with DeTEcT, which was quite a demanding task, but we hope to come back to this task in the future and analyse the application of other numerical optimisation methods.\par
For each optimisation method listed above, we have added a maximum iterations parameter, which triggers the optimisation process to stop once that number of iterations is reached. This allows us to manually control when the optimisation should stop if it does not find a satisfactory solution. By default, we define the satisfactory solution as the one that has a loss of less than $0.0001$ on the respective loss function, but of course this parameter has to be tailored to the loss function being used and the total number of tokens that exist in the economy (i.e., since the number of tokens is the unit of the loss function, except loss function like log-loss where log-tokens act as the unit). If the accuracy tolerance is set very low on purpose, we advise to set the maximum number of iterations as the hard stop to exit the optimisation problem with the ``best'' result in finite time and prevent a potentially infinite loop. It might be a good idea to set the accuracy tolerance to be very low to examine how quickly different numerical methods will converge to an infinitesimal loss (Or if they ever will), check if we can find the case where loss stagnates and doesn't go below a certain value, and to see whether the choice of a loss function impacts the optimisation procedure. In general, we can re-run backward propagation simulations multiple times with different maximum number of iterations to establish at what number of iterations we reach either a loss plateau (i.e., loss that does not decrease with further iterations and oscillates around the same value), or reach a predetermined loss level. From our experience, we recommend to start by running the same backward propagation setup with $10$, $100$ and $1,000$ maximum number of iterations as a starting point, and readjusting this parameter further from there if needed.\par
Another comment we need to add is that in our implementation we define a numerical engine that we use for the optimisation through a polymorphism, which enables us to adapt and connect any optimisation method to the rest of the implementation without performing a significant refactoring of the rest of the code base with each new implementation. The losses are computed and recorded per objective function call as it is fast and is useful for logging purposes, while different numerical methods have different number of function calls per iteration. Combined with numerical methods like Nelder-Mead, where the number of function calls per iteration is dependent on whether certain internal conditions are met, it makes it quite difficult and computationally costly to record or map the losses per iteration, while also recording losses per function call. Therefore, we keep track of losses with respect to function calls for optimisation purposes, which is the reason all the loss plots we present in this paper use number of function calls on the $x$-axis, rather than the number of iterations. This approach allows us to compare how many function calls different numerical methods make per number of iterations, and how long these simulations take in terms of runtime.\par
Following the selection of multiple numerical optimisation methods to evaluate along with different combinations of parameters for them, another configuration we suggest to experiment with is the loss function. Different loss functions may prove to be better suited for specific numerical methods or scenarios (As we will later see in our case study), so experimenting with those may add more information for future simulations.\par
Having identified the ingredients that we need to use to find an optimal set of parameters for further backward propagation simulations, we summarise the methodology with the following sequence of steps:
\begin{itemize}
	\item \textbf{[Step 1]} Creation of taxonomy of agent categories
	\item \textbf{[Step 2]} Collection and partition of wealth distribution data into data per agent category per timestamp
	\item \textbf{[Step 3]} Identification of boundary conditions data points: for forward propagation - initial wealth distribution; for backward propagation - initial and final wealth distribution, with either the largest difference in their wealth distribution or the largest time difference
	\item \textbf{[Step 4]} Creation of interactions and rotations topology, which defines permitted interaction and rotation types in the economy
	\item \textbf{[Step 5]} Definition of the dynamical system that follows from the taxonomy of agent categories and the topology of interactions and rotations
	\item \textbf{[Step 6: Forward Propagation]} Simulation of the dynamical system using the Euler method
	\item \textbf{[Step 6: Backward Propagation]} Identification of numerical optimisation methods, loss functions, accuracy tolerance, maximum number of iterations, and maximum number of function calls
	\item \textbf{[Step 7: Backward Propagation]} Simulation of the dynamical system using every combination of the parameters defined in the previous step
	\item \textbf{[Step 8: Backward Propagation]} Selection of the combination of parameters that either optimise either the accuracy or runtime, depending on the order of priority or balance between the two
\end{itemize}
This method is very general, but it should cover almost all cases when we would possibly want to optimise the parameters of our simulations for following backward propagation simulations, which is applicable for event analysis framework that we will formalise in next section, and for choosing parametrisation that can be reused for other DeTEcT-based applications.\par
\subsection{Case Study: Modelling Bitcoin Economy}
Having outlined the methodology for defining a token economy and selection of appropriate tools for its backward propagation, we now demonstrate how we would apply this methodology to model a real-world token economy in Bitcoin.\par
As described in Section \ref{Section:Scope}, the data we use is the wealth distribution between agents of the Bitcoin token economy partitioned into $10$ Gini-like buckets with a daily frequency. Therefore, the agent category taxonomy consists of $10$ buckets that define agents in different wealth bounds in the economy, with a \emph{Control Mechanism} as the 11th agent category, which represents the Bitcoin protocol, and its wealth is the unissued tokens (i.e., the difference between maximum supply and current circulation).\par
Next, we define the boundary conditions for our dynamical system. In order to find the optimal combination of numerical methods and parameters for backward propagation we use the wealth distribution recorded in January 2009 and the wealth distribution recorded in August 2025. Table \ref{Table:TechnicalConfiguration:BoundaryConditions} presents the agent categories (i.e., wealth distribution buckets) with their respective starting and final wealths respectively. The aim of a DeTEcT simulation is to find the solution of the dynamical system such that it starts at the initial wealth level and ends at the respective final wealth level for each agent category.\par
\begin{table}[ht]
\footnotesize
\begin{center}
\begin{tabular}{ |c|c|c| }
\hline
\rotatebox{90}{Agent Category} & \rotatebox{90}{Initial Wealth (BTC)} & \rotatebox{90}{Final Wealth (BTC)} \\ 
\hline
Control Mechanism & 20,999,940.98 & 1,093,216.38 \\ 
\hline
From 0 to 0.001 & 1.00 & 5,578.49 \\ 
\hline
From 0.001 to 0.01 & 1.00 & 42,483.58 \\ 
\hline
From 0.01 to 0.1 & 1.00 & 268,059.02 \\ 
\hline
From 0.1 to 1 & 1.00 & 1,068,570.95 \\ 
\hline
From 1 to 10 & 1.00 & 2,061,778.06 \\ 
\hline
From 10 to 100 & 50.00 & 4,309,046.90 \\ 
\hline
From 100 to 1,000 & 1.00 & 4,921,234.02 \\ 
\hline
From 1,000 to 10,000 & 1.00 & 4,371,132.72 \\ 
\hline
From 10,000 to 100,000 & 1.00 & 2,216,180.35 \\ 
\hline
From 100,000 to infinity & 1.00 & 642,719.49 \\ 
\hline
\end{tabular}
\end{center}
\caption{Initial (July 2009) and final (August 2025) wealth distribution states}
\label{Table:TechnicalConfiguration:BoundaryConditions}
\end{table}
With regards to interactions and rotations topology, the interactions between agents in different buckets are well defined, but the rotations prove to be difficult to justify: rotations are the movements of wealth from agent category to another due to individual agents (i.e, Bitcoin addresses in this case study) moving between the two agent categories, whereas in the Bitcoin ecosystem addresses are very rarely, if ever, reused as a security precaution (i.e., UTXO mechanism). In Bitcoin, new addresses are usually created as transactions' outputs, and the data partition we use is defined in terms of addresses. This means that old addresses are very unlikely to be reused as targets for new transactions, and therefore, are very unlikely to rotate into other Gini buckets. Hence, we assume that there are no rotations between agent categories in the economy, and set the matrix of rotations to be all $0$'s. Note that the maximum supply of Bitcoin is static and is capped at roughly $21$ million BTC. This implies that we do not need to use the dynamic supply extension of DeTEcT \cite{DeTEcTExtension}, which is an additional simplification in this case.\par
As we are planning on using the parametrisation of DeTEcT to perform event analysis with BIPs, we assume that system has static deterministic parameters, which we hypothesise should change after the introduction of a policy (i.e., when BIP is proposed). This helps us to parameterise DeTEcT in preparation for further backward propagation simulations in the later section.\par
Taking into account the settings and constraints we mentioned above, the system that describes the evolution of wealth distribution in the Bitcoin token economy according to DeTEcT is:
\begin{equation}
\label{Equation:TechnicalConfiguration:AdaptedForm}
	\frac{\Delta}{\Delta t}[\vv{F}(t)] = \frac{1}{M}\vv{F}(t)\odot[\mathcal{B}\cdot\vv{F}(t)]
\end{equation}
where $M$ is the maximum supply of Bitcoin.\par
Next, we propose the configuration of the simulation environment that we will test to determine what combination of parameters and features we should use for the most accurate backward propagation results. We find the default accuracy of $0.0001$ to be reasonable, as it is a very low accuracy for our simulations, and we would be surprised to see any numerical method to achieve it. The reason we don't set the error tolerance to a higher number that the numerical optimisation method can reach is because we cannot objectively justify a specific selection of a tolerance without being too arbitrary, so we rely on the maximum number of iterations as the hard stop to stop the simulation.\par
For our case study, wee chose multiple different maximum number of iterations to test: $100$, $500$ and $1,000$. These will allow us to compare the number of iterations that it takes for a simulation to achieve ``decent'' accuracy (Recall that it is computationally intensive to hook into the number of iterations different numerical engines make at runtime due to our polymorphism definition of a numerical optimisation method), while also giving us additional metric for the performance of the numerical method in terms of runtime. The choice of $100$, $500$ and $1,000$ maximum iterations is somewhat arbitrary. When testing gradient descent at $5,000$ and $10,000$ iterations, the simulations have failed to complete within the time period dedicated to them, and so for the purpose of this case study we will not be presenting the partial results that we obtained when running simulations at $5,000$ and $10,000$ iterations.\par
We chose two loss functions to analyse in our simulations, namely MAE (i.e., Mean Absolute Error) and MLE (i.e., Mean Log Error) that are defined as:
\begin{equation}
\begin{split}
	& MAE = \frac{1}{n}\sum^{n}_{i=1}\lvert o_{i}-p_{i} \rvert \\
	& MLE = \sqrt{\frac{1}{n}\sum^{n}_{i=1}\left[ln\left(\frac{o_{i} + 1}{p_{i} + 1}\right)\right]^{2}}
\end{split}
\end{equation}
where $o_{i}$ is the observed value (i.e., actual wealth of agent category $i$ at final time step), $p_{i}$ is the predicted value (i.e., simulated wealth of agent category $i$ at final time step), and $n$ is the number or label of an agent category (i.e., $11$ distinct agent categories in our simulations).\par
For convenience, we plot both MAE and MLE on the log-scale in order to be able to make comparisons between them (i.e., we transform MAE losses to the log-scale), and to better see big swings in losses that we may not be able to detect if we were looking at a regular loss scale. Therefore, the units that we use for the losses plotted in Appendix \ref{Appendix:NumericalMethodSimulations} are log-Bitcoins.\par
Having described the main moving parts that we stress-test and the variables we use for the simulations, we have run the simulations to produce the results presented in Appendix \ref{Appendix:NumericalMethodSimulations}, where the simulations are grouped by the combination of numerical optimisation method used along with the loss function. We now showcase how we would proceed with an analysis of a token economy model, and how to interpret the results one obtains from DeTEcT simulations.\par
\subsubsection{Numerical Optimisation Considerations}
The first item we would like to bring to attention is the absence of ``gradient descent-MAE'' combination of features. The reason for its absence is that during the initial iterations of the simulation, gradient descent could not complete numerical integration leading to the simulation crashing out every time we run it.\par
Another issue we encountered when running the case study simulations was the numerical errors in them, and specifically in Nelder-Mead-based simulations. Nelder-Mead optimisation method is deterministic by design, but the plots it produces in Appendix \ref{Appendix:NumericalMethodSimulations} for different number of maximum iterations do not map to each other. The reason for this discrepancy is the fact that Nelder-Mead method is a heuristic search method, which in the context of minimisation methods, means that it uses comparisons of function evaluations to produce a ``good-enough'' solution to optimisation problem as opposed to using a combination of Jacobian and Hessian of the objective function to crawl to the most optimal solution using the gradients and curvatures. It is a minimisation method for functions with domain in $\mathbb{R}^{n}$ and it generally works well for a lot of optimisation problems, but as the dimensionality of the domain increases Nelder-Mead method becomes less and less accurate. This happens because some of the points of the simplex, which acts as a set of points that is assumed to encompass the argmin of the objective function, produce very similar values that then have to be sorted by the loss of the function at each of those points during the algorithm. However, because of the numerical errors, the values that are close together may get sorted differently each time the simulation is re-run, which leads to different optimisation results for each simulation. This is one of the reasons why gradient-based optimisation techniques (e.g., gradient descent) are preferred for high-dimensional optimisation problems. As a slight remark we also wanted to mention that when we tested Nelder-Mead on an economy with $3$ agent categories, and $4$ interaction rates and $2$ rotation rates, Nelder-Mead has performed on par in loss optimisation with gradient descent and it was very robust in its selection of parameters (Our implementation of Nelder-Mead method has also passed multiple unit tests with low-dimensionality problems that we have set up for it).\par
\subsubsection{Analysis of Numerical Optimisation Results}
These issues aside, we have run $15$ simulations in total for all combinations of numerical methods, loss functions and maximum iteration configurations (Not accounting  for the $3$ simulations of the gradient descent and MAE combination). For each simulation we produced a log-loss plot that demonstrate the log-loss against the number of function calls, which are presented in Appendix \ref{Appendix:NumericalMethodSimulations}. Additionally, Table \ref{Table:TechnicalConfiguration:NumericalMethodsComparison} summarises some of the simulations' metadata such as the number of function calls, simulation runtime, and initial and final log-losses (All losses presented in the table have been transformed to log-scale for the presentation purposes).\par
\begin{figure}[!ht]
\centering
    \begin{subfigure}{1\linewidth}
        \includegraphics[width=\linewidth]{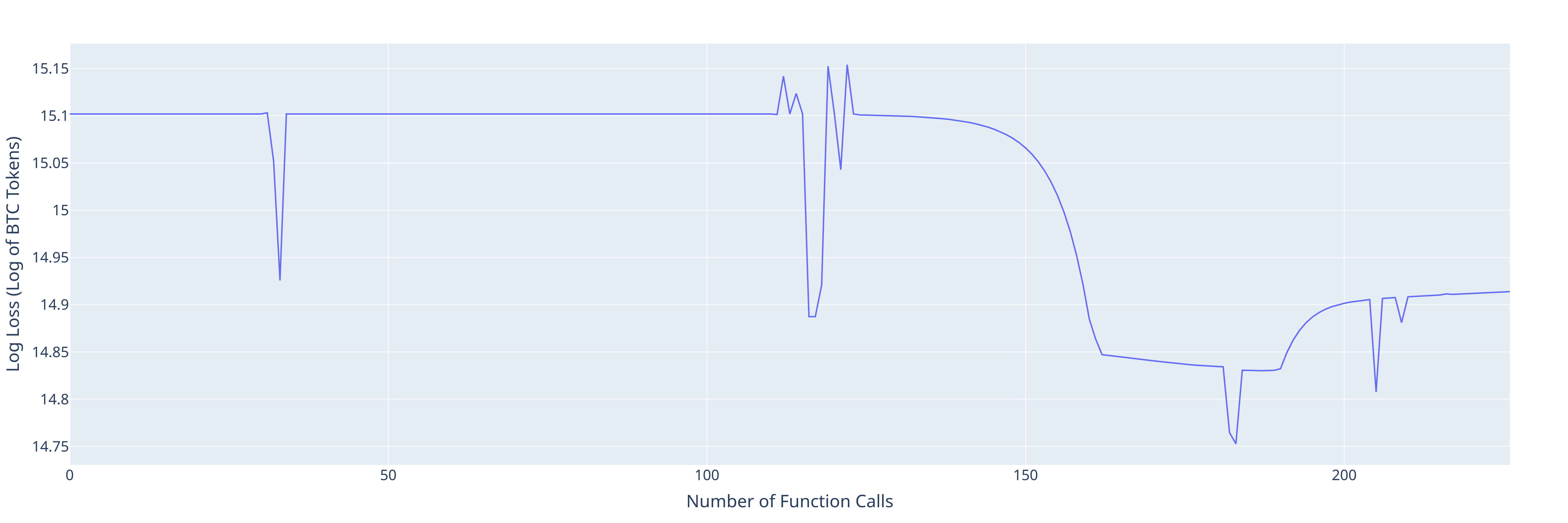}
    \caption{\footnotesize Log-losses\\(100 maximum iterations)}
    \end{subfigure}

    \begin{subfigure}{1\linewidth}
        \includegraphics[width=\linewidth]{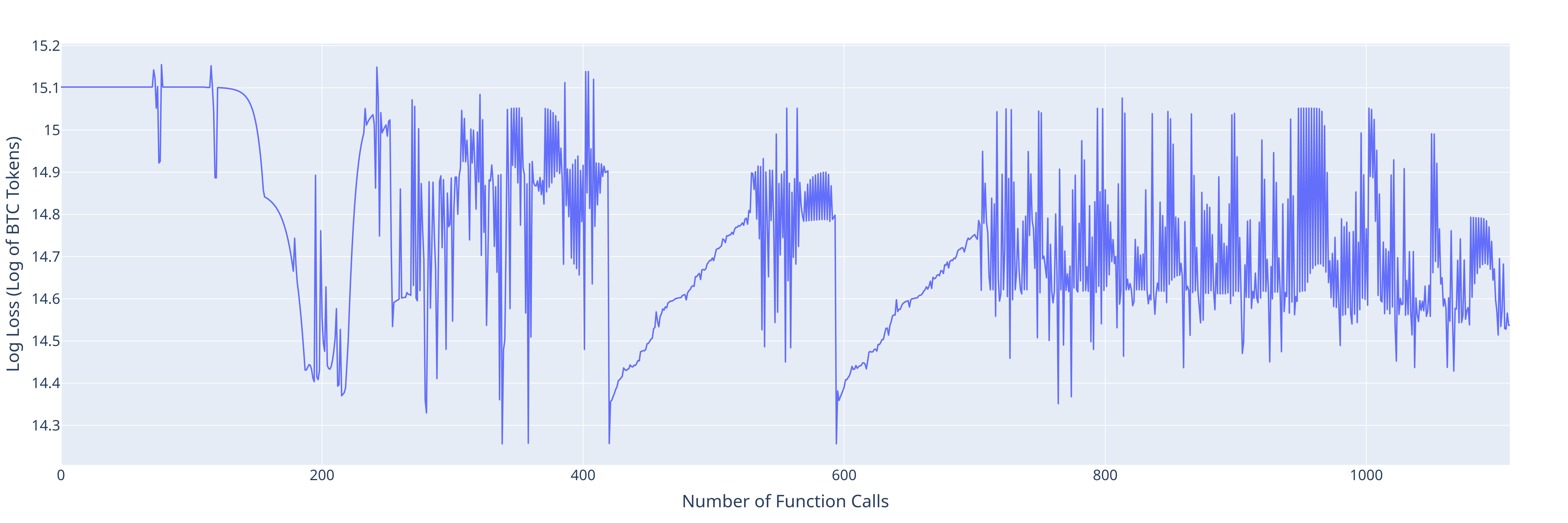}
    \caption{\footnotesize Log-losses\\(500 maximum iterations)}
    \end{subfigure}

    \begin{subfigure}{1\linewidth}
        \includegraphics[width=\linewidth]{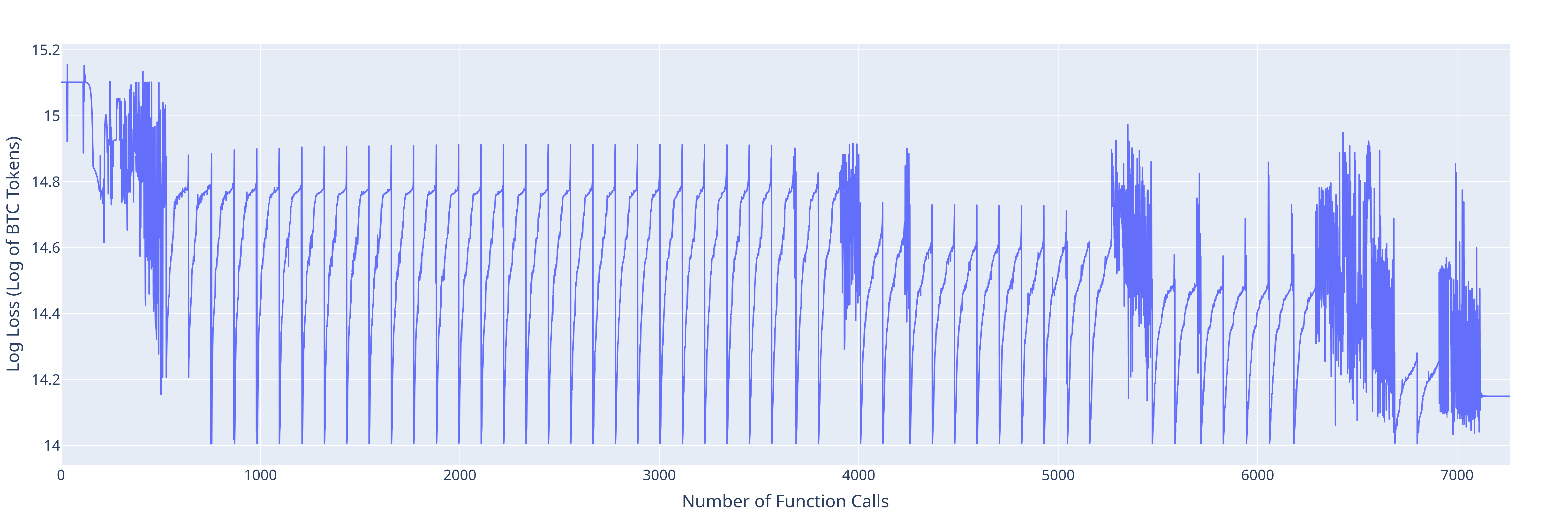}
    \caption{\footnotesize Log-losses\\(1,000 maximum iterations)}
    \end{subfigure}
\caption{Simulations of wealth distribution in Bitcoin token economy using Nelder-Mead numerical engine and MAE loss function.}
\label{Figure:NM:MAE}
\end{figure}
\begin{figure}[!ht]
\centering
    \begin{subfigure}{1\linewidth}
        \includegraphics[width=\linewidth]{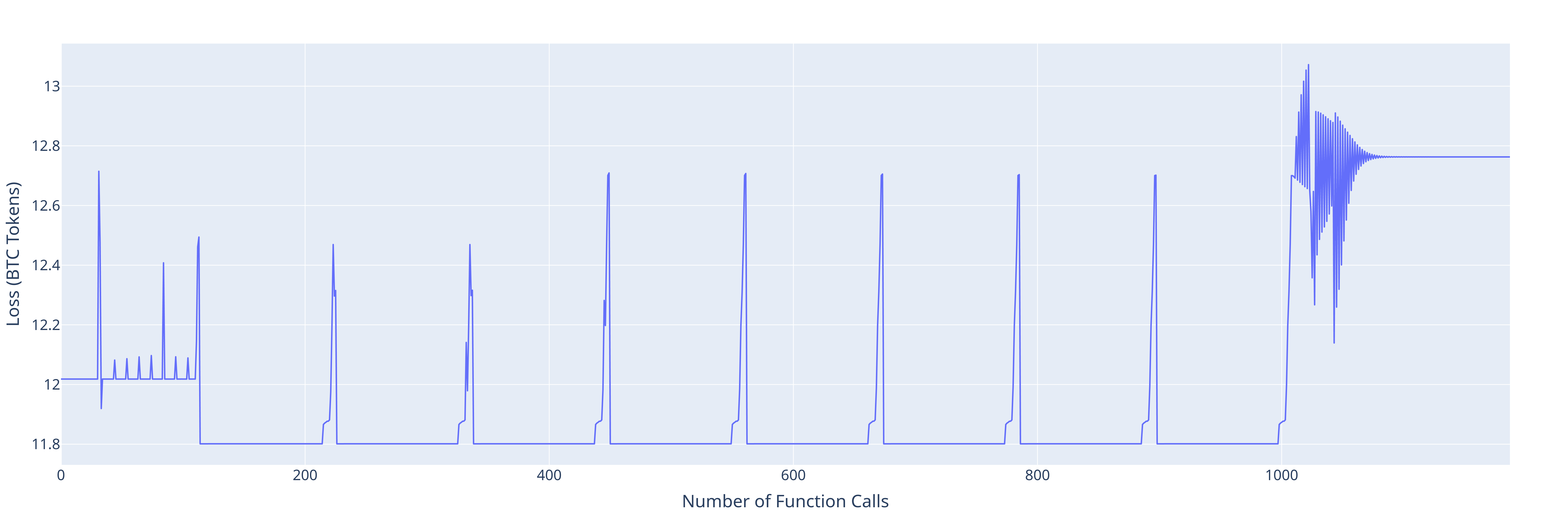}
    \caption{\footnotesize Log-losses\\(100 maximum iterations)}
    \end{subfigure}

    \begin{subfigure}{1\linewidth}
        \includegraphics[width=\linewidth]{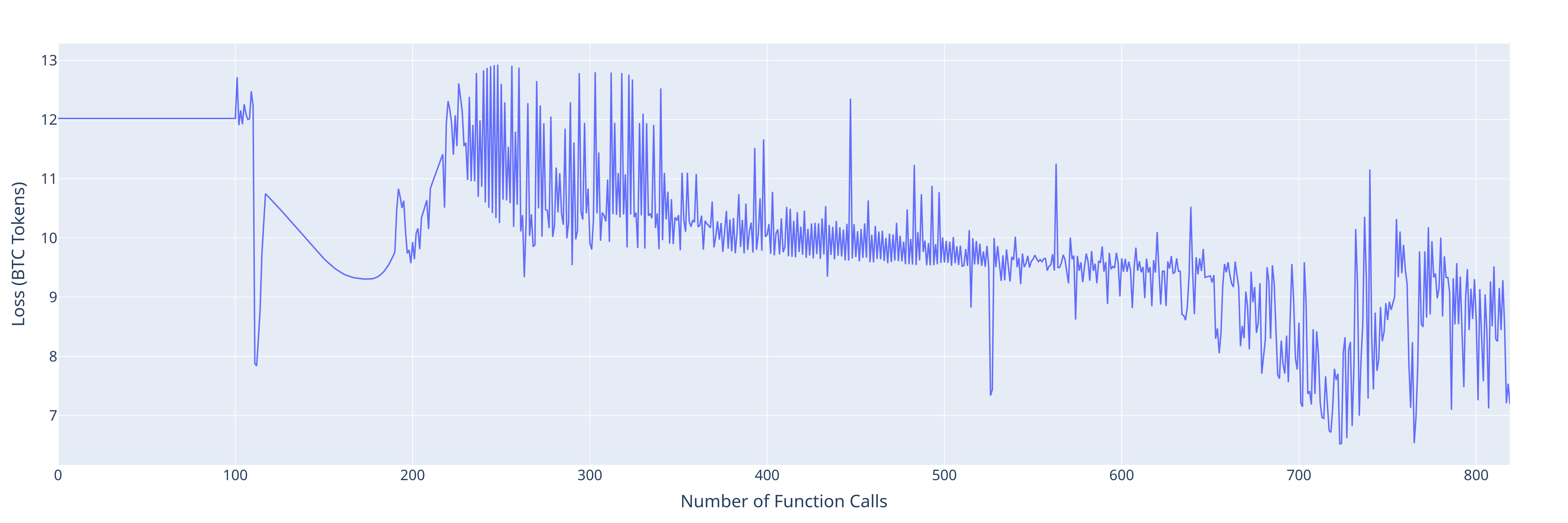}
    \caption{\footnotesize Log-losses\\(500 maximum iterations)}
    \end{subfigure}

    \begin{subfigure}{1\linewidth}
        \includegraphics[width=\linewidth]{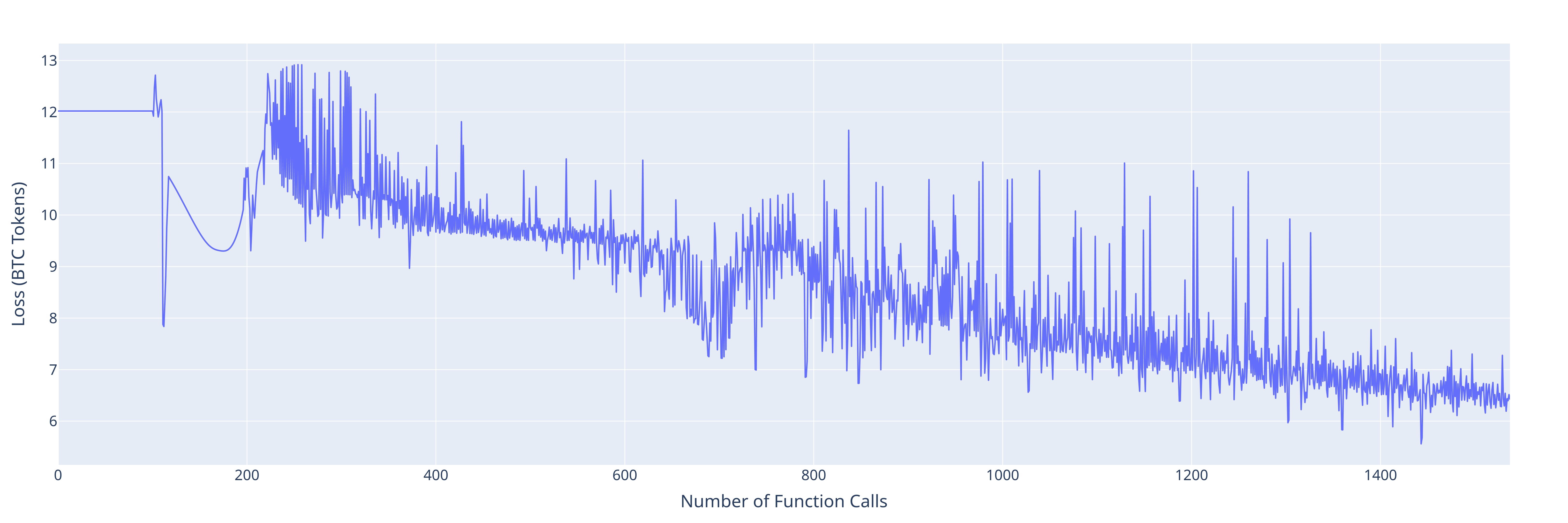}
    \caption{\footnotesize Log-losses\\(1,000 maximum iterations)}
    \end{subfigure}
\caption{Simulations of wealth distribution in Bitcoin token economy using Nelder-Mead numerical engine and MLE loss function.}
\label{Figure:NM:MLE}
\end{figure}
\begin{figure}[!ht]
\centering
    \begin{subfigure}{1\linewidth}
        \includegraphics[width=\linewidth]{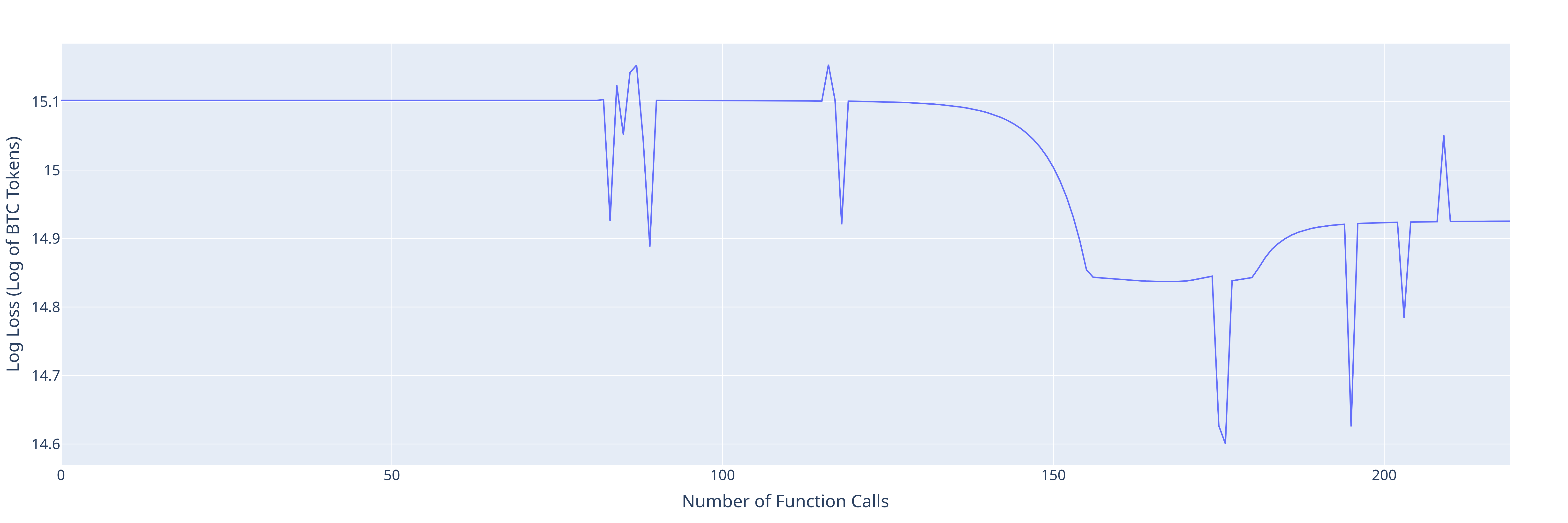}
    \caption{\footnotesize Log-losses\\(100 maximum iterations)}
    \end{subfigure}
    
    \begin{subfigure}{1\linewidth}
        \includegraphics[width=\linewidth]{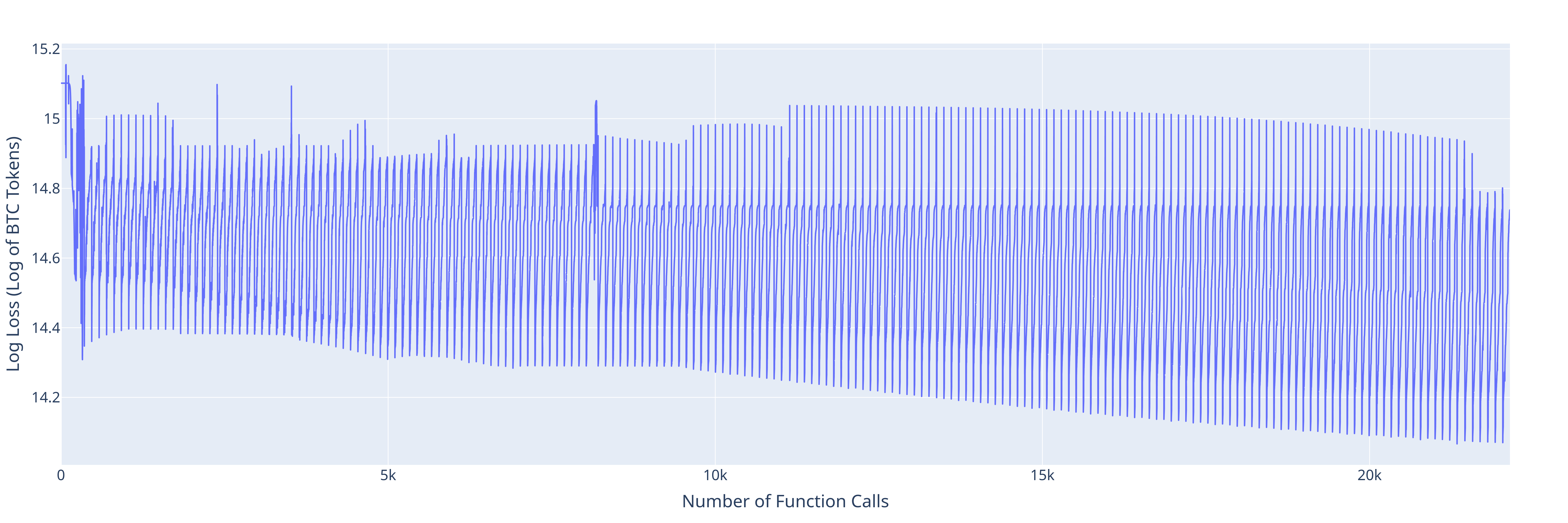}
    \caption{\footnotesize Log-losses\\(500 maximum iterations)}
    \end{subfigure}

    \begin{subfigure}{1\linewidth}
        \includegraphics[width=\linewidth]{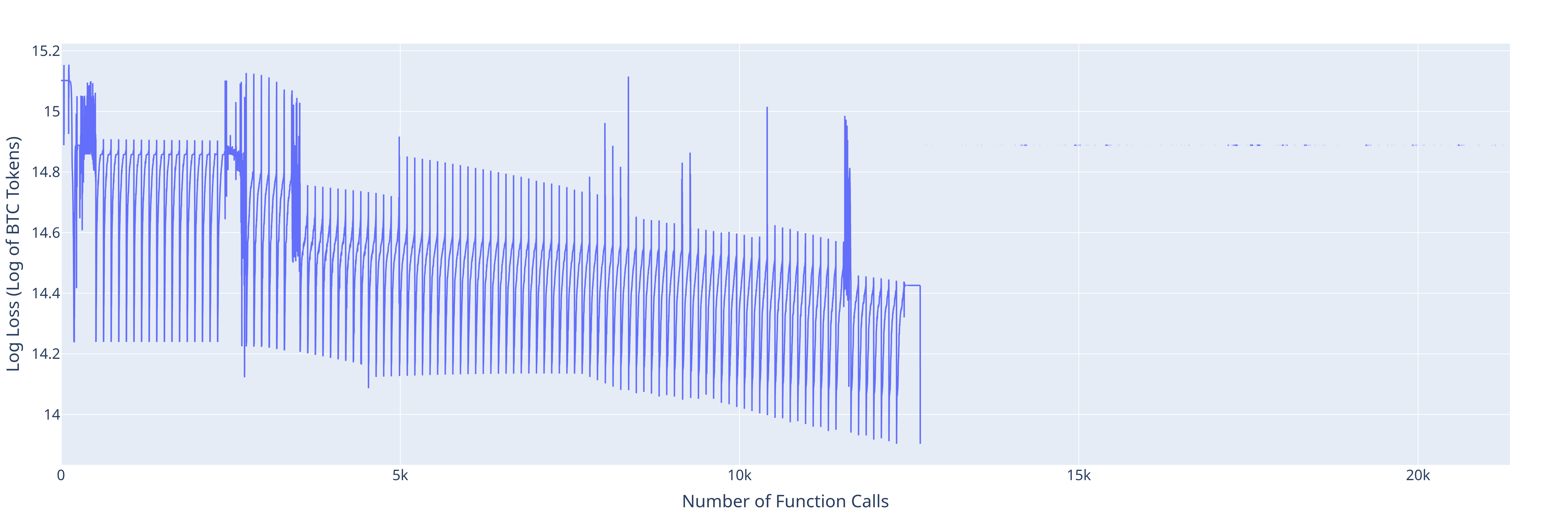}
    \caption{\footnotesize Log-losses\\(1,000 maximum iterations)}
    \end{subfigure}
\caption{Simulations of wealth distribution in Bitcoin token economy using Nelder-Mead numerical engine with random jumps and MAE loss function.}
\label{Figure:NMRJ:MAE}
\end{figure}
\begin{figure}[!ht]
\centering
    \begin{subfigure}{1\linewidth}
        \includegraphics[width=\linewidth]{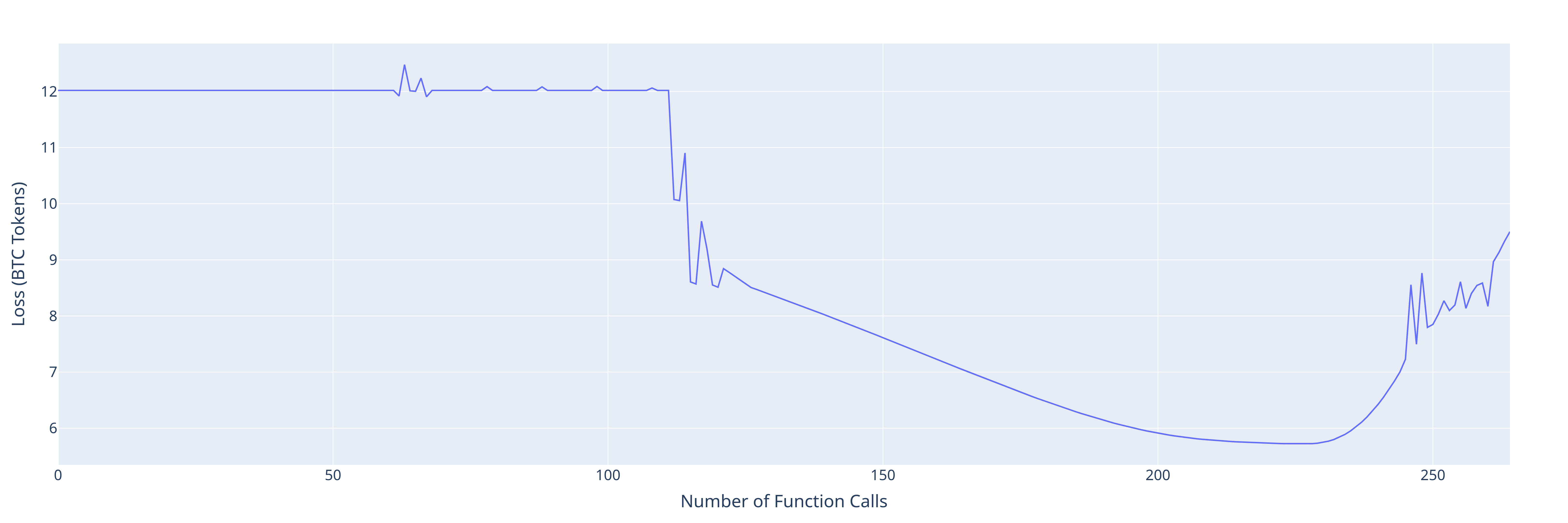}
    \caption{\footnotesize Log-losses\\(100 maximum iterations)}
    \end{subfigure}
    
    \begin{subfigure}{1\linewidth}
        \includegraphics[width=\linewidth]{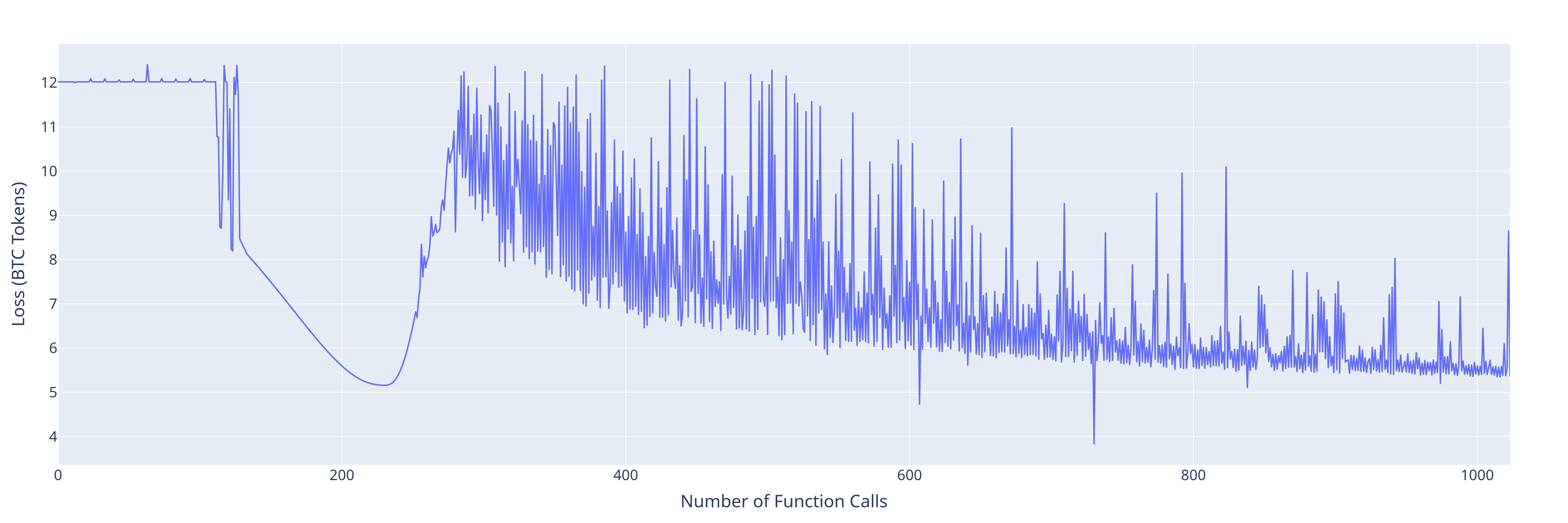}
    \caption{\footnotesize Log-losses\\(500 maximum iterations)}
    \end{subfigure}

    \begin{subfigure}{1\linewidth}
        \includegraphics[width=\linewidth]{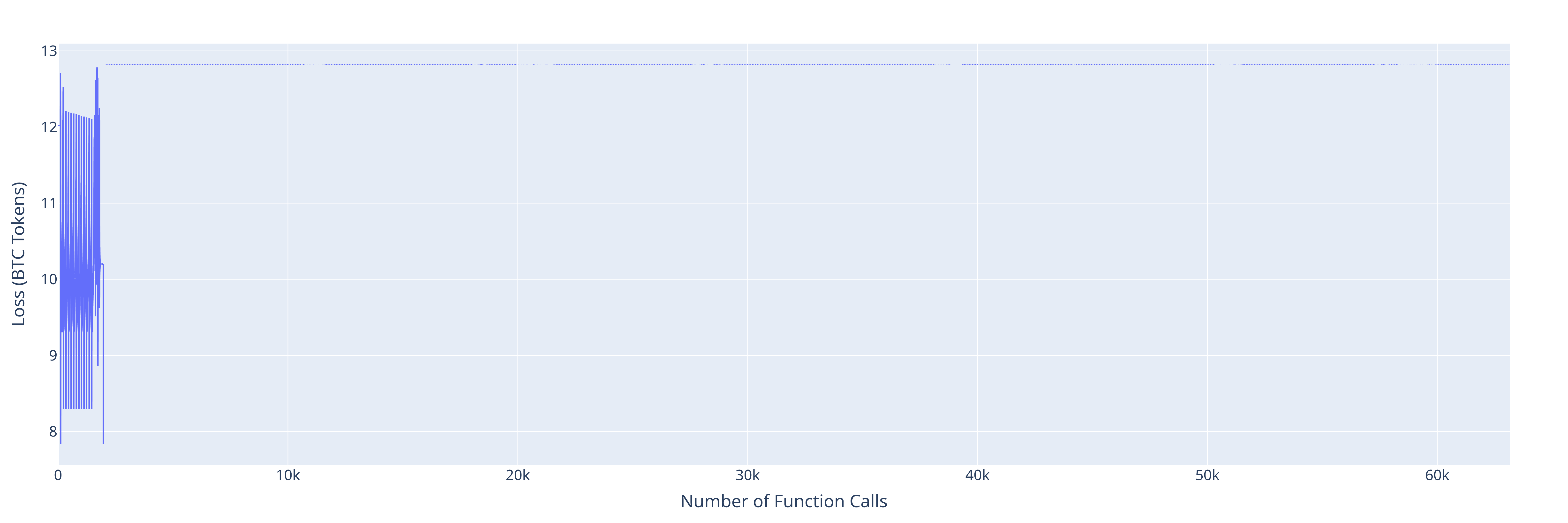}
    \caption{\footnotesize Log-losses\\(1,000 maximum iterations)}
    \end{subfigure}
\caption{Simulations of wealth distribution in Bitcoin token economy using Nelder-Mead numerical engine with random jumps and MLE loss function.}
\label{Figure:NMRJ:MLE}
\end{figure}

\begin{table}[ht]
\footnotesize
\begin{center}
\begin{tabular}{ |c|c|c|c|c|c|c| }
\hline
\rotatebox{90}{Numerical Method} & \rotatebox{90}{Loss Function} & \rotatebox{90}{Number of Iterations } & \rotatebox{90}{\begin{tabular}{@{}c@{}} Number of Objective \\ Function Calls\end{tabular}} & \rotatebox{90}{Runtime (Seconds)} & \rotatebox{90}{Initial Log-Loss} & \rotatebox{90}{Final Log-Loss} \\ 
\hline
Gradient Descent & MLE & 1,000 & 200,417 & 34,649 & 12.018266 & 0.000002 \\ 
\hline
Gradient Descent & MLE & 500 & 98,644 & 17,399 & 12.018266 & 0.000004 \\ 
\hline
Gradient Descent & MLE & 100 & 18,282 & 3,157 & 12.018266 & 0.000002 \\ 
\hline
Nelder-Mead (Random Jumps) & MLE & 1,000 & 63,161 & 11,646 & 12.018266 & 12.819492 \\ 
\hline
Nelder-Mead (Random Jumps) & MLE & 500 & 1,024 & 178 & 12.018266 & 5.378721 \\ 
\hline
Nelder-Mead (Random Jumps) & MLE & 100 & 265 & 45 & 12.018266 & 9.500035 \\ 
\hline
Nelder-Mead (Random Jumps) & MAE & 1,000 & 21,354 & 3,875 & 15.101820 & 14.888262 \\ 
\hline
Nelder-Mead (Random Jumps) & MAE & 500 & 22,146 & 3,818 & 15.101820 & 14.736274 \\ 
\hline
Nelder-Mead (Random Jumps) & MAE & 100 & 220 & 37 & 15.101820 & 14.925404 \\ 
\hline
Nelder-Mead & MLE & 1,000 & 1,538 & 268 & 12.018266 & 6.428377 \\ 
\hline
Nelder-Mead & MLE & 500 & 820 & 144 & 12.018266 & 7.193402 \\ 
\hline
Nelder-Mead & MLE & 100 & 1,188 & 205 & 12.018266 & 12.762866 \\ 
\hline
Nelder-Mead & MAE & 1,000 & 7,266 & 1,245 & 15.101820 & 14.149515 \\ 
\hline
Nelder-Mead & MAE & 500 & 1,111 & 191 & 15.101820 & 14.537016 \\ 
\hline
Nelder-Mead & MAE & 100 & 227 & 39 & 15.101820 & 14.913762 \\ 
\hline
\end{tabular}
\end{center}
\caption{Numerical optimisation methods comparison}
\label{Table:TechnicalConfiguration:NumericalMethodsComparison}
\end{table}
Examining the plots, we see that Nelder-Mead performs quite poorly when it comes to optimising the dynamical system in equation \ref{Equation:TechnicalConfiguration:AdaptedForm} as we see from Figures \ref{Figure:NM:MAE}, \ref{Figure:NM:MLE}, \ref{Figure:NMRJ:MAE} and \ref{Figure:NMRJ:MLE}, which again can be explained by the high dimensionality of our optimisation problem as we have $11$ agent categories, which leads to $55$ interaction rates that we need to find  (Recall that the matrix of interaction rates is antisymmetric and its diagonal values are $0$, which is why the dimensionality of the problem is not $121$). We can see that the log-loss dips below $6$ only on a handful of occasions, and we finish the backward propagation under $6$ Only once, when we combine Nelder-Mead with random jumps, MLE loss function and set the maximum number of iterations to $500$ - then we get the log-loss of $5.38$. However, when we increase the number of iterations to $1,000$ we immediately see an increase in loss. The potential reason for this is that the numerical solver got stuck and made a random jump to reset its position, and it could not find a better solution since.\par
Another pattern we see in the data is that when combined with MLE, Nelder-Mead method tends to perform slightly better. We go from the log-loss of $12.76$ at $100$ iterations to log-loss of $6.43$ at $1,000$ iterations, while we only see a marginal increase in accuracy when comparing Nelder-Mead with MAE combination - decrease in log-loss from $14.91$ at $100$ iterations to $14.15$ at $1,000$ iterations. It seems that the combination of Nelder-Mead with MLE is the only case where the number of iterations mattered.\par
Moving onto gradient descent, the results are a lot more interesting. Gradient descent has not only managed to achieve good accuracy, but it almost hit the final boundary condition spot-on, which is a remarkable precision with the highest final log-loss being only $0.000004$. Additionally, gradient descent method proved to be very robust with regards to the numerical errors in the parameters (i.e., interaction rates) as irrespective of the maximum number of interactions we obtained the same parameters up to $5$ significant figures. It seems gradient descent has not been significantly impacted by the number of iterations either, just like the case of Nelder-Mead with random jumps. However, the log-loss plots for gradient descent are quite messy and offer little information due to the function calls that were used to compute the gradient in different directions, which led to large jumps in the loss on the plots.\par
The main issue we came across when using gradient descent is that due to the large number of objective function calls, it was extremely slow at the same number of iterations as Nelder-Mead. For comparison, it took Nelder-Mead only $4.5$ minutes to find a solution with MLE loss function, while it took just over $9.5$ hours and nearly $200,000$ more function calls for gradient descent to complete the backward propagation with otherwise the same configuration.\par
The choice of loss function proves to be quite important as well, as in our case MAE did not work with gradient descent due to issues with numerical integration, and additionally, when we swapped it for MLE - Nelder-Mead has improved in terms of performance and reduced its runtime. We cannot draw any conclusions about the loss functions from the data produced by Nelder-Mead method with random jumps as random jumps skew the loss function  performance analysis.\par
Having followed the methodology outlined in Section \ref{Subsection:ModellingTokenEconomy}, we have defined the agent category taxonomy and topology of interactions and rotations, which enabled us to constrain the general form of dynamical system from equation \ref{Equation:TechnicalConfiguration:GeneralSystem} to obtain equation \ref{Equation:TechnicalConfiguration:AdaptedForm} that defines Bitcoin token economy from perspective of Gini-like agent categories. We then proposed a selection of the numerical engines with different combination of parameters, and loss functions, which we used to run backward propagation simulations of the token economy model we defined. A combination of gradient descent numerical method with MLE loss function has proven to be the most accurate, while it seems that running it at even $100$ maximum iterations is wasteful, since it converges to $0$ log-loss very fast, as seen from the plots. Therefore, for future backward propagation simulations of Bitcoin token economy that is defined with the same agent categories, we will use gradient descent with MLE and a maximum number of iterations of $10$.\par
\subsection{Summary}
In this section, we proposed a generalised methodology for creating models of economies using DeTEcT, and selecting appropriate methods and parameters to run backward propagation simulations. This is useful for configuring DeTEcT simulation environment to run different scenarios and for other applications that are build on top of DeTEcT (e.g., in the next section we will use DeTEcT backward propagation simulations to help us formalise event analysis framework).\par
To demonstrate the application of this methodology, we presented a case study in which we tested the application of different numerical methods with respect to DeTEcT and we found that Nelder-Mead, despite being the fastest method we tested, drops in  accuracy, which can potentially be explained by the high dimensionality of the optimisation problem that we tasked it with (Albeit working really well in simple low dimensionality unit tests we have set for it). The addition of random jumps to Nelder-Mead method has not improved its accuracy, while also being computationally intensive. For a simplified token economy model without rotations, gradient descent was the best performing numerical method (MAE integration-related issues aside) and delivered really good results in the form of accuracy and robustness to numerical errors, but at an expense of a high runtime and large number of function calls.

\section{Event Analysis with DeTEcT}
\label{Section:EventAnalysis}
In previous section we have formalised a way to define a token economy model and select the optimal configuration for backward propagation simulations. In this section we build an event analysis framework that uses DeTEcT simulation results as an input to define quantitative metrics that measure the significance of the impact of events on an economy and its compartments, as well as providing tools for qualitative analysis of economic activity. As in the previous section, we first define the generalised methodology for performing the analysis, and then we apply it in a case study where we use Bitcoin token economy that we defined in Section \ref{Section:TechnicalConfiguration}, along with BIPs as the events whose significance on different compartments of the economy we measure. We validate this case study against the empirical causality analysis that we have conducted earlier in a separate research paper \cite{BTCImpactsOfPolicies}.\par
\subsection{General Methodology: Event Analysis Framework}
\label{Subsection:EventAnalysis:GeneralMethod}
By event analysis framework we mean that we want to define a formal way to study how an event has impacted an economy, which requires a metric or number of metrics that quantify this impact. Additionally, we want to understand how significant this impact is, as a metric in the absence of context may produce a misleading result. We also want to understand what agent categories (i.e., compartments) of the economy have been impacted as we can imagine the case where the economy is not significantly impacted by an event, yet one of its agent categories is.\par
As we have seen in the previous section, one of the prerequisites of a DeTEcT token economy model is the partitioning of agents in the economy into agent categories, which together with the definition of interaction and rotation topology enables us to represent the movement of wealth in the economy using interaction and rotation rates. These parameters control different ways the wealth can move between agent categories, and they act as a quantitative metric. By running a backward propagation simulation we obtain the interaction and rotation rates that quantify the behaviour of agent categories in respect to one another.\par
Putting all of these ingredients together, we propose to run two backward propagation simulations, before and after an event, and then computing the difference in the interaction and rotation rates to determine how an event has impacted the dynamics of the token economy after it has been introduced.\par
Below we formalise this procedure in distinct steps for convenience, which are then summarised at the end of the section.\par
\textbf{[Step 1]} We first configure our simulation environment for backward propagation simulations, which we described earlier in Section \ref{Section:TechnicalConfiguration}. This also includes the creation of agent category taxonomy, interactions and rotations topology, and the construction of the dynamical system (Note that for this application we must have a model with features compatible with backward propagation).\par
\textbf{[Step 2]} Next, the reasonable question would be ``How long before and how long after should the simulations extend?'', which by default can be the start and end of the available data itself. Each event will have a \emph{window}, or \emph{horizon}, when it impacts an economy, and the default approach may encompass this window, but it is a sub-optimal solution as it may use too much data, which leads to performance issues and simply errors due to the noise and other disturbances in the data. The better approach is to perform a Granger-causality test on the wealth distribution data using event or events as the independent variable that is tested to ``cause'' the change in wealth distribution data \cite{BTCImpactsOfPolicies}. This allows us to identify the largest significant lagged term of the event signal time series, which tells us what is the longest period of time that it takes for the event to impact the wealth distribution data, and by extension, the economy.\par
\textbf{[Step 3]} We then run two backward propagation simulations, where the \emph{Before} simulation starts at the timestamp of the event minus the event impact window, and runs until the timestamp of the event, while the \emph{After} simulation runs from the timestamp of the event and goes up to the timestamp of the event plus the event impact window (e.g., if event is in August and event impact window is two months, the \emph{Before} simulation should run from June to August, and \emph{After} simulation - from August to October). The idea  is that these simulations show how the economy model parameters have changed after the event took place.\par
\textbf{[Step 4]} At this stage we have obtained matrices of interaction and rotation rates, which we can examine for patterns in the data that may provide further context to the simulations. This part of analysis is more qualitative, but there are some generic patterns in the interaction and rotation rates that we can look our for. We can examine what compartments of the economy have the largest and smallest interaction rates (Note that the magnitude of the interaction rate does not directly correspond to the size of wealth movement, but it does signify at what rate an agent category would transact its wealth away in comparison to its other interaction rates, or potentially hoard the wealth if the interaction rates are very low), which signify what are the main counterparties of this agent category's transactions or what other agent categories it relies on for transactions. Additionally, the definition of a token economy under DeTEcT requires the presence of a \emph{Control Mechanism} that represents a token distribution mechanism or protocol, so in the context of token economies we would like to examine the interaction rates of the \emph{Control Mechanism} (Note that by definition it should not have any rotations defined to and from it) to provide us with better context of how other economic participants interact with it, and whether it tells us anything about token issuance in the economy. Doing a comparative analysis may also prove useful, where we look at the matrices of interaction and rotation rates for \emph{Before} and \emph{After} simulations to compare how the patterns we see in them change after the event.\par
\textbf{[Step 5]} Next, we need to provide the context to compare the change in the interaction and rotation rates to, as without it we won't be able to determine what changes in the interaction and rotation rates are significant. To achieve this, we run \emph{Before} and \emph{After} simulations with the same event impact window, but for every possible data point in the wealth distribution dataset that is available (Because of the event impact window, we won't be able to run the simulations for some of the earliest and latest data points as they won't have enough data before and after themselves respectively). If the dataset is too large to run simulations on all of it, it may might make more sense to select a smaller, but still substantially large subset to run these contextual simulations on. This step gives us a perspective of how much do the interaction and rotation rates of the economy change after a ``regular'' day (i.e., a day with or without some event attached to it).\par
\textbf{[Steps 6-9]} With the \emph{Before}, \emph{After}, and contextual simulations all complete, we can now formalise the test for the impact of an event on a token economy. Recall that the matrix $\mathcal{B}$ of interaction rates is antisymmetric as the influx of wealth into some arbitrary agent category $A_{1}$ from $A_{2}$ is the outflux of wealth from $A_{2}$ to $A_{1}$ (i.e., just a change of the sign). To examine the difference between interaction rate matrices of \emph{Before} and \emph{After} simulations, we can sum the total absolute difference between the matrices and get a scalar metric that represents the total difference, but we can also sum the difference per agent category in each of the simulations (Due to the antisymmetric property we can sum either the difference of columns or rows of the interaction rates matrices) and obtain a metric for how the interactions of different agent categories with each other have changed according to DeTEcT. Note that from perspective of economics, the interaction rates are well defined, but their difference doesn't make much sense, so we use this technique purely as the indicator of change.\par
As for the rotation rates, the matrix of rotation rates $\Gamma$ in not antisymmetric and the only constraint it has is that the values in the columns have to sum-up to $0$. We can measure the total difference in the same way as with interaction rate matrices, but for the total difference per agent category we will get two statistics, since if we sum the difference of columns we get the difference in rotations into the agent category, while  if we sum the difference of rows we get the difference in rotations out of the respective agent category.\par
We formalise these statistics as:
\begin{equation}
	\Delta\mathcal{B}_{total} = \sum^{n}_{i=1}\sum^{n}_{j=1}\lvert\beta_{after, ij} - \beta_{before, ij}\rvert
\end{equation}
\begin{equation}
	\Delta\mathcal{B}_{ac, j} = \sum^{n}_{i=1}\lvert\beta_{after, ij} - \beta_{before, ij}\rvert
\end{equation}
\begin{equation}
	\Delta\Gamma_{total} = \sum^{n}_{i=1}\sum^{n}_{j=1}\lvert\gamma_{after, ij} - \gamma_{before, ij}\rvert
\end{equation}
\begin{equation}
	\Delta\Gamma_{ac, in, i} = \sum^{n}_{j=1}\lvert\gamma_{after, ij} - \gamma_{before, ij}\rvert
\end{equation}
\begin{equation}
	\Delta\Gamma_{ac, out, j} = \sum^{n}_{i=1}\lvert\gamma_{after, ij} - \gamma_{before, ij}\rvert
\end{equation}
where $n$ is the total number of agent categories, $\beta_{ij}$ and $\gamma_{ij}$ are respectively the interaction and rotation rates between agent categories labelled $i$ and $j$, $\Delta\mathcal{B}_{total}$ is the total difference in the interaction rates, $\Delta\mathcal{B}_{ac, j}$ is the total difference in the interaction rates of the agent category labelled $j$, $\Delta\Gamma_{total}$ is the total difference in the rotation rates, $\Delta\Gamma_{ac, in, i}$ is the total difference in the rotation rates incoming to the agent category labelled $i$, and $\Delta\Gamma_{ac, out, j}$ is the total  difference in the rotation rates outgoing from the agent category labelled $j$.\par
Having defined our metrics, the next step is to compute the contextual metrics, where we first apply the metrics above to every single pair of \emph{Before} and \emph{After} contextual simulations that we have run, and then we find the mean and standard deviation for each of these metrics. This tells us the expected change in interaction and rotation rates at a specific time, regardless of whether any significant event has occurred on it, as well as the standard deviation. Using the means and standard deviations we can construct an interval for each metric to understand whether the value of the metric falls within $1$ standard deviation or not, and if it does not we assume that the value of the metric is significant.\par
One of the main reasons we decided to go this route for measuring the significance in the change of the rates is because in general the model of a token economy defined by equation \ref{Equation:TechnicalConfiguration:GeneralSystem} does not contain a stochastic term (e.g., white noise) that would add randomness to the simulations, so it is impossible to derive a metric to use in a statistical test to measure the significance of the change in rates using statistical significance tools like t-test (Since the standard deviation of the model will be $0$ due to no random  parts). Different parameter extension for DeTEcT that we could use in conjunction with the backward propagation simulations behave this way, so we cannot use the statistical significance approach to measure significance.\par
Finally, we can distil this general methodology to the following steps:
\begin{itemize}
	\item \textbf{[Step 1]} Configure the economy model and simulation environment as described in Section \ref{Section:TechnicalConfiguration}
	\item \textbf{[Step 2]} Selection of an event impact window (Can be achieved by selecting the lag of the highest significant lagged term in the Granger-causality analysis of the event signal time series)
	\item \textbf{[Step 3]} Run \emph{Before} and \emph{After} simulations around the event timestamp
	\item \textbf{[Step 4]} Perform a qualitative analysis of the interaction and rotation rate matrices, some of the points of interest for which can be summarised as, but not limited to:
	\begin{itemize}
		\item Largest and smallest interaction and rotation rates in the economy may explain what are the largest and smallest points of wealth turnover in the economy
		\item Largest and smallest interaction and rotation rates per agent category may explain what other agent categories the agent category depends on and transacts with the most, and whether it hoards its wealth
		\item For token economies, comparing interaction rates of \emph{Control Mechanism} may illuminate patterns in token distribution and the dependence of agent categories on token issuance by the protocol
		\item Comparative analysis of interaction and rotation rates for \emph{Before} and \emph{After} simulations may explain what exact wealth redistribution routes have been impacted by the event
	\end{itemize}
	\item \textbf{[Step 5]} Run contextual simulations for all available data with the same event impact window
	\item \textbf{[Step 6]} Computation of the difference metrics of the \emph{Before} and \emph{After} simulations' interaction and rotation rate matrices
	\item \textbf{[Step 7]} Computation of the same metrics for all contextual simulations, along with their means and standard deviations
	\item \textbf{[Step 8]} Calculation of the significance bounds for each metric using the means and standard deviations
	\item \textbf{[Step 9]} If the metric of the event is outside of its respective bounds, the event has significantly impacted that compartment of the economy (Or economy as a whole if it is the total interaction or rotation rate difference)
\end{itemize}
Now that we have defined all the necessary tools, we proceed to demonstrate their application in our Bitcoin case study.\par
\subsection{Case Study: Application of Event Analysis Framework to Bitcoin}
The aim of this case study is to demonstrate how the event analysis framework that we defined can be applied to examine the impacts of events on real-world token economies. To achieve this, we set up a series of simulations where for each \emph{Major Economy-Related BIP} (i.e., BIPs 32, 42, 50, 141, 341) we run two simulations: first - $6$ months leading up to the BIP introduction, and second - $6$ months after the BIP introduction. Since we will be using the same Bitcoin wealth distribution dataset as before, the token economy is still defined by equation \ref{Equation:TechnicalConfiguration:AdaptedForm}, gradient descent with $10$ maximum iterations and MLE is our configuration for the backward propagation simulations, and $6$ months is event impact window that we have found in our prior empirical analysis \cite{BTCImpactsOfPolicies}.\par
With the configuration complete, we have run $10$ simulations (i.e., \emph{Before} and \emph{After} simulations for $5$ BIPs) with their interaction rate matrices presented in Appendix \ref{Appendix:MajorEconomicBIPsSimulationsIRMatrices}. Note that all BIPs that we used are set more than $6$ months apart, which means that no simulation was overlapping with the next, making  the results ``clean'' from the impacts of another \emph{Major Economy-Related BIP}. We can now perform a quantitative analysis of the interaction rates that we have obtained from the \emph{Before} and \emph{After} simulations to demonstrate how DeTEcT can be used to provide more context for the study of a token economy.\par
The interaction rate matrices in Appendix \ref{Appendix:MajorEconomicBIPsSimulationsIRMatrices} demonstrate a solution to the optimisation problem described above, and it is important to note that there is no uniqueness of solution, so it might be possible to find a different set of interaction rates for these simulations. For all of the simulations described in this section we used a matrix of $0$'s as the initial guess for the interaction rates, so the gradient descent has found the set of interaction rates that produced the smallest loss around the origin in our parameter space. The reason we use $0$ as the origin and our initial guess is because the interaction rate of $0$ indicates no interaction, which we assume to be the default.\par
In the tables in the appendix, the individual interaction rates can be interpreted as the rate at which the wealth moves from the agent category in the respective column to the agent category in the respective row. The negative sign means that the wealth moves from the agent category in the respective row to the agent category in the respective column. Also note that the diagonal interaction rates are all $0$ because the interactions of individual agents in a specific agent category will leave the aggregate wealth of that agent category unchanged, so they are left as the default values.\par
Having described how to read the interaction rates tables, we now proceed to analysing the individual interaction matrices. We start by looking at the general patterns that can be attributed to the simulations' parameters.\par
First, note that pretty much every interaction rate attributed to \emph{Control Mechanism} is set to outgoing, meaning that \emph{Control Mechanism} spends tokens in transactions with other agent categories. This makes sense from the perspective of Bitcoin token economy as \emph{Control Mechanism} represents the protocol that issues tokens. The reason for some inflows of wealth into the \emph{Control Mechanism} is that the simulations were not constrained to prevent that interaction (i.e., protocol receiving tokens), which is something we would like to add to our implementation in the future.\par
Another pattern we identify in the interaction rate matrices is that the poorer the bucket the more other buckets tend to move wealth into it through interactions; only after BIP 341 is there a pattern reversal where the poorest bucket starts spending tokens when interacting with wealthier buckets. This might be explained by the design of the Bitcoin transaction system - when an address is used to transact, its wealth covers the transaction value, while the leftover goes to a new address (i.e., the UTXO mechanism). This means that in the Bitcoin ecosystem we often see bigger addresses being broken down into a bunch of smaller ones when processing a transaction. The possible reason why the inverse of this operation is rare (i.e., multiple addresses uniting into one address) is because there can be many addresses contained in the same wallet so there is no practical reason to unite them into one address before a new transaction (In fact that unification may cost a transaction fee to perform, which further disinsentivises this behaviour). Additionally, Bitcoin is commonly used as a speculative financial instrument due to its utility in portfolio diversification (Bitcoin is considered an alternative asset) and regulatory arbitrage, which deflates the value of Bitcoin for internal transactions (i.e., transaction inside Bitcoin economy). This means that over time the nominal transaction values for goods and services priced in Bitcoin fall over time, further reinforcing the deflation loop. This may also partially explain the pattern of wealth migrating towards the poorer addresses.\par
The largest interaction rates usually involve at least one counterparty from the richer buckets or \emph{Control Mechanism}, which tends to have the largest interaction rates by magnitude. The smallest interaction rates by magnitude tend to be between two agent categories from the following set: \emph{From 0 to 0.001}, \emph{From 0.001 to 0.01}, \emph{From 0.01 to 0.1} and \emph{From 0.1 to 1}. From economics perspective, this means that due to the design of the system, agent categories are incentivised to interact with \emph{Control Mechanism} and in Bitcoin it can be explained by mining, which becomes the main source of wealth generation and circulation in the token economy. This in turn may result in a ``waterfall'' of tokens down the wealth distribution buckets, but it comes at an expense of economic activity as agent categories prioritise interactions with \emph{Control Mechanism} and not interactions with other agent categories. It creates a form of interactions centralisation in the economy, where agents become dependent on the distribution of wealth from the protocol, and the reduction in interactions with it by other agent categories may lead to hoarding (i.e., other agent categories relying on distribution of tokens via mining; the drop in mining rewards may cause agent categories to hoard wealth as they start to earn less), further deflating the value of Bitcoin as a currency.\par
Next step in our framework is to run the contextual simulations. For these we have used the $6$ month event impact window, and  we have run the contextual simulations for every day from 11-02-2012 until 19-01-2020, which we have used to compute the mean and standard deviation for the total difference of interaction rate matrices, as well as the aggregate difference in interaction rates per agent category. These results are presented in Appendix \ref{Appendix:MajorEconomicBIPsSimulationsDailyChanges}. Recall that the \emph{Lower Bound} and \emph{Upper Bound} columns from the table in the appendix are defined as the mean value minus and plus the standard deviation respectively.\par
In Section \ref{Subsection:EventAnalysis:GeneralMethod}, we introduced the metrics to measure the change in the interaction and rotation rates due to an event. Specifically for interaction rates, we have introduced $\Delta\mathcal{B}_{total}$ to measure the total difference in the interaction rates, and $\Delta\mathcal{B}_{ac,j}$ to measure the total difference in the interaction rates of an agent category labelled $j$. We now apply this metrics to the interaction rate matrices in Appendix \ref{Appendix:MajorEconomicBIPsSimulationsIRMatrices} to obtain Tables \ref{Table:EventAnalysis:Total} and \ref{Table:EventAnalysis:PerAgentCategory}. Table \ref{Table:EventAnalysis:Total} presents $\Delta\mathcal{B}_{total}$ for every BIP (i.e., for every event), while Table \ref{Table:EventAnalysis:PerAgentCategory} presents $\Delta\mathcal{B}_{ac,j}$ for the same BIPs. Note that by definition of these metrics, we sum absolute values, so these metrics do not provide us with any indication of the direction of overall changes (i.e., they don't provide any context for whether interaction rates have increased or decreased on average). In addition to these tables, we computed the mean and standard deviation of these metrics from our contextual simulations in order to provide the significance context to these metrics. The table with means, standard deviations, and computed significance bounds is presented in Appendix \ref{Appendix:MajorEconomicBIPsSimulationsDailyChanges}.\par
\begin{table}[ht]
\footnotesize
\begin{center}
\begin{tabular}{ |c|c|c|c|c|c|c| }
\hline
 & \rotatebox{90}{BIP 32} & \rotatebox{90}{BIP 42} & \rotatebox{90}{BIP 50} & \rotatebox{90}{BIP 141} & \rotatebox{90}{BIP 341} & \rotatebox{90}{Total} \\
\hline
Total Difference & 0.084916 & 0.092249 & 0.041913 & 0.078505 & 0.062476 & 0.360058 \\
\hline
\end{tabular}
\caption{$\Delta\mathcal{B}_{total}$ for each BIP (i.e., total difference in the interaction rates due to the respective BIP)}
\label{Table:EventAnalysis:Total}
\end{center}
\end{table}
\begin{table}[ht]
\footnotesize
\begin{center}
\begin{tabular}{ |c|c|c|c|c|c|c| }
\hline
 & \rotatebox{90}{BIP 32} & \rotatebox{90}{BIP 42} & \rotatebox{90}{BIP 50} & \rotatebox{90}{BIP 141} & \rotatebox{90}{BIP 341} & \rotatebox{90}{Total} \\
\hline
Control Mechanism & \cellcolor{green!25} $0.030944^{\ast}$ & 0.024727 & 0.012637 & 0.015958 & 0.006922 & 0.091187 \\
\hline
From 0 to 0.001 & 0.004847 & 0.005156 & 0.001211 & 0.005656 & 0.004556 & 0.021427 \\
\hline
From 0.001 to 0.01 & 0.003846 & 0.003224 & 0.001975 & 0.005207 & 0.001781 & 0.016033 \\
\hline
From 0.01 to 0.1 & 0.003097 & 0.005157 & 0.004836 & 0.000846 & 0.001732 & 0.015667 \\
\hline
From 0.1 to 1 & 0.001462 & \cellcolor{green!25} $0.007287^{\ast}$ & 0.004320 & \cellcolor{green!25} $0.008605^{\ast}$ & 0.002768 & 0.024442 \\
\hline
From 1 to 10 & 0.001285 & 0.004543 & 0.003615 & \cellcolor{green!25} $0.008289^{\ast}$ & 0.002904 & 0.020636 \\
\hline
From 10 to 100 & 0.003826 & 0.004202 & 0.003319 & 0.007967 & 0.005025 & 0.024338 \\
\hline
From 100 to 1,000 & 0.007308 & 0.006407 & 0.002937 & 0.006917 & 0.006001 & 0.029571 \\
\hline
From 1,000 to 10,000 & 0.003964 & 0.008788 & \cellcolor{green!25} $0.002994^{\ast}$ & 0.007246 & 0.007712 & 0.030704 \\
\hline
From 10,000 to 100,000 & 0.003256 & 0.005762 & 0.003267 & 0.005376 & 0.005680 & 0.023342 \\
\hline
From 100,000 to infinity & 0.021080 & 0.016998 & 0.000801 & 0.006438 & 0.017394 & 0.062711 \\
\hline
\end{tabular}
\caption{$\Delta\mathcal{B}_{ac,j}$ for each BIP (i.e., total difference in the interaction rates due to the respective BIP per agent category; green cells with asterisk indicate significant results)}
\label{Table:EventAnalysis:PerAgentCategory}
\end{center}
\end{table}
We can now analyse the metrics and their context to demonstrate how these can be used to draw conclusions from this event analysis framework. The first observation we make is that the largest difference in interaction rates after the introduction of a BIP was due to BIP 42, and the smallest difference was due to BIP 50, which is expected as BIP 42 proposed significant changes in the supply mechanism, while BIP 50 was a temporary chain fork and was a one-off event that should not impact agent categories long-term as much as a major protocol update like BIP 42.\par
Additionally, we use the table in Appendix \ref{Appendix:MajorEconomicBIPsSimulationsDailyChanges} to check for significance of total difference of any of these BIPs when compared to their expected value. We note that the range for significance for $\Delta\mathcal{B}_{total}$ is $(0.026826, 0.132846)$, and all of the values from Table \ref{Table:EventAnalysis:Total} lie within that range, which implies that these BIPs have not impacted the Bitcoin economy as a whole in a significant way according to our significance testing procedure. That, however, does not mean that these BIPs do not significantly impact different compartments of the economy.\par
In Table \ref{Table:EventAnalysis:PerAgentCategory} we highlighted the cells with significant results (i.e., cells whose $\Delta\mathcal{B}_{ac,j}$ metric lies outside the bounds, making them significant). We see that BIP 32 has significantly impacted \emph{Control Mechanism}, BIP 42 - \emph{From 0.1 to 1}, BIP 50 - \emph{From 1,000 to 10,000}, and BIP 141 has impacted significantly both \emph{From 0.1 to 1} and \emph{From 1 to 10} agent categories. This event analysis framework helps us to see in more detail the differences made by BIPs on specific compartments of the economy, which compliments the empirical causality analysis we have performed \cite{BTCImpactsOfPolicies}, explaining why these BIPs have a causally significant impact on the Bitcoin token economy.\par
If we are to study the nominal changes in the interaction rates without the added significance context, we see that the two agent categories with the highest interaction rate differences are \emph{Control Mechanism} and \emph{From 100,000 to infinity} with the difference of $0.091187$ and $0.062711$ respectively. The agent categories with the smallest interaction rate differences are \emph{From 0.001 to 0.01} and \emph{From 0.01 to 0.1} with the total difference being $0.016033$ and $0.015667$ respectively.\par
We see that the interaction rates of poorer agent categories on average experienced smaller change to their interaction rates after the introduction of \emph{Major Economy-Related BIPs} compared to the wealthier agent categories. However, when we look at the metric scores that are significant, we see that the results we obtained line up well with the causality analysis we performed earlier \cite{BTCImpactsOfPolicies}, where \emph{From 0.1 to 1} has been significantly impacted by these BIPs according to the simple and full Granger causality tests at $6$, $10$ and $12$ months, as well as the \emph{From 1 to 10} and \emph{From 1,000 to 10,000} being significantly impacted as also seen from the causality tests. Additionally, we note that the poorest and the wealthiest agent categories haven't been impacted, which also lines up with our causal analysis findings.\par
Another detail we would like to highlight in reference to our previous study is that we categorised BIPs 42 and 141 as being the \emph{Monetary-Like BIPs}, while BIPs 32 and 341 we considered to be \emph{Purely Tokenomic BIPs} \cite{BTCImpactsOfPolicies} (Note that BIP 50 was not categorised as a stand alone policy as it was a response to the fork). From Table \ref{Table:EventAnalysis:PerAgentCategory} we see that both \emph{Monetary-Like BIPs} have impacted the wealth buckets, while \emph{Purely Tokenomic BIPs} collectively only impacted the \emph{Control Mechanism}. Based on this result, we hypothesise that \emph{Monetary-Like BIPs} tend to impact mid-wealth wealth distribution buckets in Bitcoin, while \emph{Purely Tokenomic BIPs} seem not to affect any of the ``standard'' Gini-like wealth distribution buckets and they rather impact \emph{Control Mechanism} as the representation of the Bitcoin protocol in our model, but we leave this for future research.\par
As a last remark, we must mention that the interaction rates we presented here are the result of using gradient descent optimisation, and despite us trying to use a larger maximum number of iterations (i.e., $100$ and $500$ maximum iterations), we still receive the same interactions rates to $5$ significant figures. However, this does not indicate that these are the most optimal interaction rates and it can be the case that using a different starting point and configuration of parameters can yield a completely different set of interaction rates that optimise the loss function equally well. We analysed the interaction rates we found to see if they make sense from perspective of economics and our framework, which they do, but there still may exist better fits for this model of wealth distribution in Bitcoin.\par
\subsection{Summary}
In this section, we proposed an event analysis framework that uses DeTEcT's backward propagation simulations to test for the significance of impacts of events on an economy. This framework is capable of demonstrating that an event has changed the dynamics in the economy with significance, and it is capable of highlighting what compartments of the economy are the most affected by this event. The utility of this framework is in analysing the impacts of events and policies on the economy, while potentially explaining what compartments they target and how effective they are at targeting these specific compartments.\par
To demonstrate the application of the framework we used Bitcoin token economy as a case study, where we analysed the impact of BIPs (i.e., Bitcoin endogenous policies) on different Gini-like wealth distribution buckets. The quantitative analysis revealed the effects of UTXO mechanism with potentially a large degree of interactions of poorer buckets being the unspent transaction outputs, while also demonstrating the significant dependence of agent categories on \emph{Control Mechanism}, which created a centralisation of interactions as agent categories seemed to heavily rely on mining to receive tokens with major changes to wealth dynamics being always ``funded'' by \emph{Control Mechanism}.\par
The results of our event analysis metrics have been validated by our causal analysis of the empirical data, as there seems to be a strong overlap between the agent categories that have been significantly impacted by the \emph{Major Economy-Related BIPs}. This demonstrates that our event analysis framework can be used to study what compartments of the economy the event impacts significantly, which is a useful tool for understanding what parts of the economy the policies are actually impacting in a significant way. Additionally, we hypothesise that the placement of a policy in a policy taxonomy \cite{EconomicPolicyTaxonomy} may predict what compartments of the economy it will significantly impact, with \emph{Purely Tokenomic BIPs} seemingly not impacting the Gini-like wealth distribution buckets and rather impacting \emph{Control Mechanism}, while \emph{Monetary-Like BIPs} having the significant impact on some of the agent categories, which shows that monetary-like policies can change wealth distribution in Bitcoin economy.\par

\section{Future Work}
Throughout this paper, we picked up a few points where we would like to improve future implementations of DeTEcT and its applications. First, we would like to adopt and test DeTEcT with more numerical optimisation methods such as Powell's method, SLSQP and L-BFGS. Addition of these numerical methods allows us to test their performance when applied to DeTEcT, while expansing the customisation options available to our frameworks. Also, it might be the case that the choice of a numerical method may depend on the topology of the economy at hand, which is a research question that we would like to examine in the future.\par
In Section \ref{Section:EventAnalysis} we proposed a technique for analysing the impact of BIPs, as endogenous policies, on the Bitcoin token economy. In the future, we would like to introduce a metric for measuring the direction of change in the interaction rates in \emph{Before} and \emph{After} simulations. This would allow us to better understand the direction-of-action for the event in addition to our proposed methodology for determining the compartments it affects.\par
We would also like to add a feature to our implementation that will allow us to set different constraints to different parts of the simulation. In our analysis we came across some interaction rates that lead to the inflow of wealth into \emph{Control Mechanism}, which was not well defined for the Bitcoin token economy in our case study. Therefore, we would like to build a mechanism for adding constraints for interaction and rotation rates, which are applied at the start of the simulation and don't result in a significant rise in runtime.\par
Lastly, we would like to explore the dependency of the placement of an atomic policy in the policy taxonomy on what compartments of the economy this atomic policy affects in a significant way. That is to say we would like to understand whether knowing the impact of an atomic policy can be used to forecast the impacts of atomic policies around it in the economic policy taxonomy. However, this research will require a lot of additional data and experimentation with many structurally-different economies, so we leave it for a future project.

\section{Conclusion}
The objective of this paper was to formalise the setup of a token economy model, configuration of the backward propagation simulations, and propose an event analysis framework that can be used to measure and study the impacts of events on an economy. Aside from presenting the formal frameworks, this paper is also meant to demonstrate how to apply the defined frameworks to a real-world case study.\par
We have broken down the main considerations that go into using DeTEcT to create a model of a token economy, where the partition of economy's participants into agent categories is an important step along with processing of wealth distribution data. We have explained how to select an appropriate configurations to run backward propagation simulations with DeTEcT, including different loss functions, numerical optimisation methods and their parametrisation.\par
Having formalised the configuration of the model and its simulation environment, we introduced a method for analysing the impact of an event on an economy as a whole, and on specific agent categories therein defined. We proposed to run one simulation before and one after the event with the objective to compare the interaction and rotation rates of the two simulations in order to understand what the event has changed about the economy. To measure the changes we introduced a number of metrics that help us to quantify the changes, while with an inclusion of a contextual analysis step, we added a way to check whether the change is significant, or it could be attributed to the regular noise of the data.\par
The frameworks and tools described in this paper have been applied to create a model of Bitcoin token economy given the partitioning of agents into Gini-like wealth distribution buckets. We have shown that Nelder-Mead and gradient descent can be applied to running DeTEcT's backward propagation simulations, and that the choice between different numerical optimisation methods may present a dilemma between accuracy and runtime. Then, using the selected settings, we have performed an analysis of impacts of Bitcoin's endogenous policies (i.e., BIPs) on the token economy and its agent categories using the methodology that we formalised. The results we have obtained using our analysis framework are validated by the empirical analysis of the same data that we have performed prior, which demonstrates that DeTEcT can be used to perform analysis of events, while also adding some contextual details, such as showing what compartments of the economy have been affected by an event and quantifying how much they were affected.\par
In conclusion, we demonstrated how to apply DeTEcT to modelling token economies, while we also showcased its potential in economic analysis. It helped us quantify the changes made by policy introductions and fundamental events, and explain what compartments of a token economy are impacted by what policies using the formal analysis methodology we proposed. We also learned multiple lessons about the structure of token economies through the case study of Bitcoin as well as studying the structure of Bitcoin itself, which lays the ground for further quantitative research on policy implementation in token economies.

\footnotesize 

\newpage
\appendix
\appendixpage
\section{Numerical Methods Simulations}
\label{Appendix:NumericalMethodSimulations}

\newpage

\newpage

\newpage

\newpage
\begin{figure}[!ht]
\centering
    \begin{subfigure}{1\linewidth}
        \includegraphics[width=\linewidth]{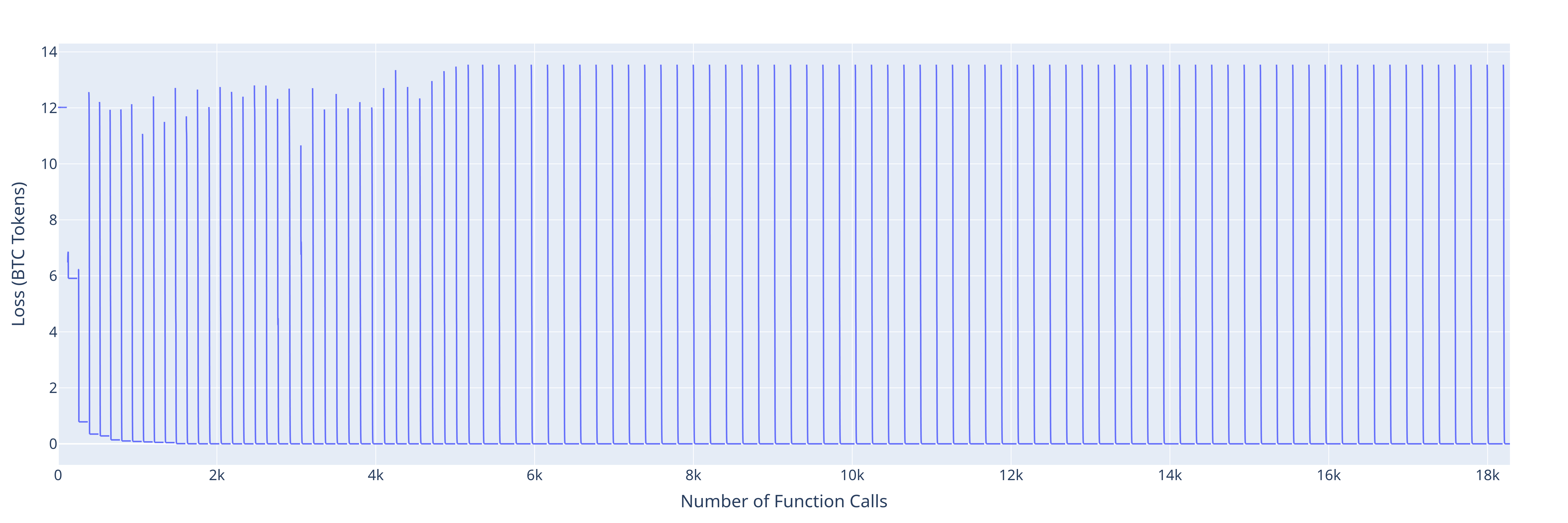}
    \caption{\footnotesize Log-losses\\(100 maximum iterations)}
    \end{subfigure}
    
    \begin{subfigure}{1\linewidth}
        \includegraphics[width=\linewidth]{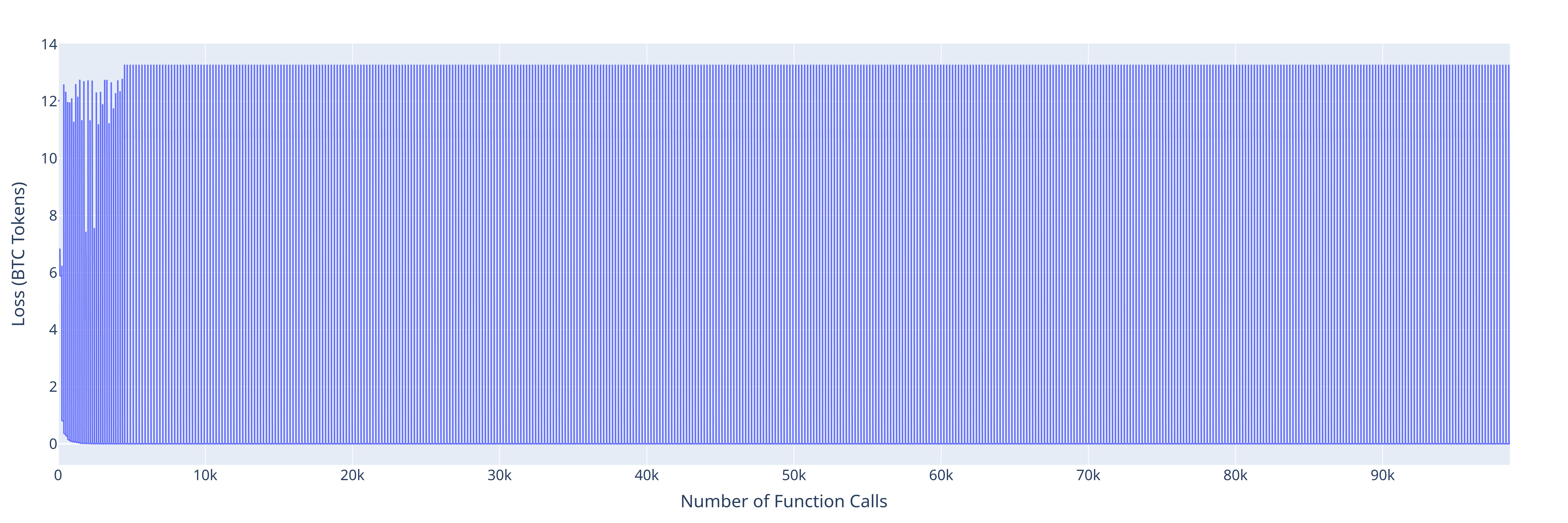}
    \caption{\footnotesize Log-losses\\(500 maximum iterations)}
    \end{subfigure}

    \begin{subfigure}{1\linewidth}
        \includegraphics[width=\linewidth]{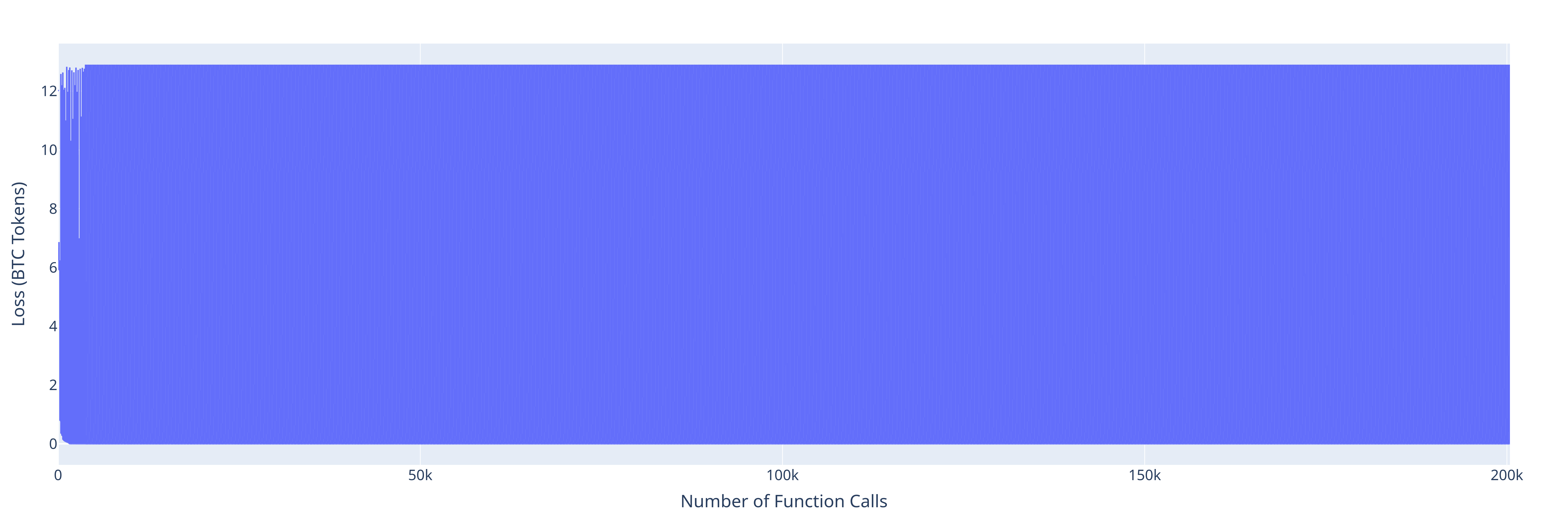}
    \caption{\footnotesize Log-losses\\(1,000 maximum iterations)}
    \end{subfigure}
\caption{Simulations of wealth distribution in Bitcoin token economy using Gradient Descent numerical engine with and MLE loss function.}
\label{Figure:GD:MLE}
\end{figure}

\newpage
\section{BIP Simulations: Interaction Rate Matrices}
\label{Appendix:MajorEconomicBIPsSimulationsIRMatrices}
\begin{table}[hbt!]
\scriptsize
\begin{center}
\rotatebox{90}{
\begin{tabular}{ |c|c|c|c|c|c|c|c|c|c|c|c| }
\hline
 & \rotatebox{270}{Control Mechanism} & \rotatebox{270}{From 0 to 0.001} & \rotatebox{270}{From 0.001 to 0.01} & \rotatebox{270}{From 0.01 to 0.1} & \rotatebox{270}{From 0.1 to 1} & \rotatebox{270}{From 1 to 10} & \rotatebox{270}{From 10 to 100} & \rotatebox{270}{From 100 to 1,000} & \rotatebox{270}{From 1,000 to 10,000} & \rotatebox{270}{From 10,000 to 100,000} & \rotatebox{270}{From 100,000 to infinity} \\
\hline
\multicolumn{12}{|c|}{\textbf{BIP 32 - Before}} \\
\hline
Control Mechanism & 0.000000 & -0.006200 & -0.009090 & -0.001000 & -0.002463 & -0.001630 & -0.001971 & -0.004986 & -0.002006 & -0.000900 & 0.006283 \\ 
\hline
From 0 to 0.001 & 0.006200 & 0.000000 & 0.000000 & 0.000001 & 0.000005 & 0.000053 & 0.000435 & 0.000423 & 0.000648 & 0.000548 & 0.000155 \\ 
\hline
From 0.001 to 0.01 & 0.009090 & -0.000000 & 0.000000 & 0.000001 & 0.000007 & 0.000073 & 0.000595 & 0.000528 & 0.000887 & 0.000774 & 0.000263 \\ 
\hline
From 0.01 to 0.1 & 0.001000 & -0.000001 & -0.000001 & 0.000000 & 0.000000 & 0.000008 & 0.000065 & 0.000057 & 0.000097 & 0.000085 & 0.000030 \\ 
\hline
From 0.1 to 1 & 0.002463 & -0.000005 & -0.000007 & -0.000000 & 0.000000 & 0.000018 & 0.000159 & 0.000138 & 0.000238 & 0.000208 & 0.000076 \\ 
\hline
From 1 to 10 & 0.001630 & -0.000053 & -0.000073 & -0.000008 & -0.000018 & 0.000000 & 0.000090 & 0.000054 & 0.000142 & 0.000131 & 0.000097 \\ 
\hline
From 10 to 100 & 0.001971 & -0.000435 & -0.000595 & -0.000065 & -0.000159 & -0.000090 & 0.000000 & -0.000210 & 0.000061 & 0.000108 & 0.000465 \\ 
\hline
From 100 to 1,000 & 0.004986 & -0.000423 & -0.000528 & -0.000057 & -0.000138 & -0.000054 & 0.000210 & 0.000000 & 0.000367 & 0.000369 & 0.000504 \\ 
\hline
From 1,000 to 10,000 & 0.002006 & -0.000648 & -0.000887 & -0.000097 & -0.000238 & -0.000142 & -0.000061 & -0.000367 & 0.000000 & 0.000082 & 0.000667 \\ 
\hline
From 10,000 to 100,000 & 0.000900 & -0.000548 & -0.000774 & -0.000085 & -0.000208 & -0.000131 & -0.000108 & -0.000369 & -0.000082 & 0.000000 & 0.000558 \\ 
\hline
From 100,000 to infinity & -0.006283 & -0.000155 & -0.000263 & -0.000030 & -0.000076 & -0.000097 & -0.000465 & -0.000504 & -0.000667 & -0.000558 & 0.000000 \\
\hline
\multicolumn{12}{|c|}{\textbf{BIP 32 - After}} \\
\hline
Control Mechanism & 0.000000 & -0.008997 & -0.005993 & -0.003091 & -0.001301 & -0.000784 & -0.000552 & -0.001081 & -0.002316 & -0.000476 & -0.008610 \\ 
\hline
From 0 to 0.001 & 0.008997 & 0.000000 & 0.000000 & 0.000002 & 0.000010 & 0.000097 & 0.000819 & 0.001037 & 0.001256 & 0.000936 & 0.000160 \\ 
\hline
From 0.001 to 0.01 & 0.005993 & -0.000000 & 0.000000 & 0.000001 & 0.000006 & 0.000061 & 0.000518 & 0.000641 & 0.000730 & 0.000595 & 0.000052 \\ 
\hline
From 0.01 to 0.1 & 0.003091 & -0.000002 & -0.000001 & 0.000000 & 0.000003 & 0.000031 & 0.000266 & 0.000329 & 0.000374 & 0.000306 & 0.000024 \\ 
\hline
From 0.1 to 1 & 0.001301 & -0.000010 & -0.000006 & -0.000003 & 0.000000 & 0.000012 & 0.000111 & 0.000137 & 0.000155 & 0.000128 & 0.000002 \\ 
\hline
From 1 to 10 & 0.000784 & -0.000097 & -0.000061 & -0.000031 & -0.000012 & 0.000000 & 0.000062 & 0.000072 & 0.000071 & 0.000073 & -0.000080 \\ 
\hline
From 10 to 100 & 0.000552 & -0.000819 & -0.000518 & -0.000266 & -0.000111 & -0.000062 & 0.000000 & -0.000035 & -0.000134 & 0.000013 & -0.000738 \\ 
\hline
From 100 to 1,000 & 0.001081 & -0.001037 & -0.000641 & -0.000329 & -0.000137 & -0.000072 & 0.000035 & 0.000000 & -0.000117 & 0.000056 & -0.000909 \\ 
\hline
From 1,000 to 10,000 & 0.002316 & -0.001256 & -0.000730 & -0.000374 & -0.000155 & -0.000071 & 0.000134 & 0.000117 & 0.000000 & 0.000173 & -0.001023 \\ 
\hline
From 10,000 to 100,000 & 0.000476 & -0.000936 & -0.000595 & -0.000306 & -0.000128 & -0.000073 & -0.000013 & -0.000056 & -0.000173 & 0.000000 & -0.000849 \\ 
\hline
From 100,000 to infinity & 0.008610 & -0.000160 & -0.000052 & -0.000024 & -0.000002 & 0.000080 & 0.000738 & 0.000909 & 0.001023 & 0.000849 & 0.000000 \\ 
\hline
\end{tabular}
}
\caption{BIP 32: Interaction rates matrices}
\end{center}
\end{table}

\begin{table}[hbt!]
\scriptsize
\begin{center}
\rotatebox{90}{
\begin{tabular}{ |c|c|c|c|c|c|c|c|c|c|c|c| }
\hline
 & \rotatebox{270}{Control Mechanism} & \rotatebox{270}{From 0 to 0.001} & \rotatebox{270}{From 0.001 to 0.01} & \rotatebox{270}{From 0.01 to 0.1} & \rotatebox{270}{From 0.1 to 1} & \rotatebox{270}{From 1 to 10} & \rotatebox{270}{From 10 to 100} & \rotatebox{270}{From 100 to 1,000} & \rotatebox{270}{From 1,000 to 10,000} & \rotatebox{270}{From 10,000 to 100,000} & \rotatebox{270}{From 100,000 to infinity} \\ 
\hline
\multicolumn{12}{|c|}{\textbf{BIP 42 - Before}} \\
\hline
Control Mechanism & 0.000000 & -0.007066 & -0.004866 & -0.005056 & -0.005034 & -0.003610 & -0.001275 & -0.001525 & 0.000328 & -0.000768 & -0.002423 \\ 
\hline
From 0 to 0.001 & 0.007066 & 0.000000 & 0.000000 & 0.000005 & 0.000034 & 0.000227 & 0.001103 & 0.001676 & 0.002134 & 0.001394 & 0.000071 \\ 
\hline
From 0.001 to 0.01 & 0.004866 & -0.000000 & 0.000000 & 0.000003 & 0.000023 & 0.000155 & 0.000758 & 0.001151 & 0.001470 & 0.000958 & 0.000049 \\ 
\hline
From 0.01 to 0.1 & 0.005056 & -0.000005 & -0.000003 & 0.000000 & 0.000020 & 0.000158 & 0.000786 & 0.001194 & 0.001527 & 0.000995 & 0.000049 \\ 
\hline
From 0.1 to 1 & 0.005034 & -0.000034 & -0.000023 & -0.000020 & 0.000000 & 0.000143 & 0.000778 & 0.001183 & 0.001522 & 0.000988 & 0.000039 \\ 
\hline
From 1 to 10 & 0.003610 & -0.000227 & -0.000155 & -0.000158 & -0.000143 & 0.000000 & 0.000521 & 0.000805 & 0.001101 & 0.000686 & -0.000041 \\ 
\hline
From 10 to 100 & 0.001275 & -0.001103 & -0.000758 & -0.000786 & -0.000778 & -0.000521 & 0.000000 & 0.000064 & 0.000436 & 0.000132 & -0.000364 \\ 
\hline
From 100 to 1,000 & 0.001525 & -0.001676 & -0.001151 & -0.001194 & -0.001183 & -0.000805 & -0.000064 & 0.000000 & 0.000538 & 0.000119 & -0.000557 \\ 
\hline
From 1,000 to 10,000 & -0.000328 & -0.002134 & -0.001470 & -0.001527 & -0.001522 & -0.001101 & -0.000436 & -0.000538 & 0.000000 & -0.000297 & -0.000735 \\ 
\hline
From 10,000 to 100,000 & 0.000768 & -0.001394 & -0.000958 & -0.000995 & -0.000988 & -0.000686 & -0.000132 & -0.000119 & 0.000297 & 0.000000 & -0.000469 \\ 
\hline
From 100,000 to infinity & 0.002423 & -0.000071 & -0.000049 & -0.000049 & -0.000039 & 0.000041 & 0.000364 & 0.000557 & 0.000735 & 0.000469 & 0.000000 \\ 
\hline
\multicolumn{12}{|c|}{\textbf{BIP 42 - After}} \\
\hline
Control Mechanism & 0.000000 & -0.004057 & -0.002948 & -0.002207 & -0.001209 & -0.001436 & -0.000812 & -0.000758 & -0.000617 & 0.000144 & -0.010290 \\ 
\hline
From 0 to 0.001 & 0.004057 & 0.000000 & 0.000001 & 0.000006 & 0.000038 & 0.000214 & 0.000786 & 0.001220 & 0.001254 & 0.000933 & 0.000056 \\ 
\hline
From 0.001 to 0.01 & 0.002948 & -0.000001 & 0.000000 & 0.000004 & 0.000027 & 0.000155 & 0.000571 & 0.000886 & 0.000911 & 0.000678 & 0.000039 \\ 
\hline
From 0.01 to 0.1 & 0.002207 & -0.000006 & -0.000004 & 0.000000 & 0.000019 & 0.000114 & 0.000426 & 0.000661 & 0.000680 & 0.000507 & 0.000016 \\ 
\hline
From 0.1 to 1 & 0.001209 & -0.000038 & -0.000027 & -0.000019 & 0.000000 & 0.000050 & 0.000227 & 0.000356 & 0.000368 & 0.000280 & -0.000079 \\ 
\hline
From 1 to 10 & 0.001436 & -0.000214 & -0.000155 & -0.000114 & -0.000050 & 0.000000 & 0.000235 & 0.000391 & 0.000411 & 0.000338 & -0.000522 \\ 
\hline
From 10 to 100 & 0.000812 & -0.000786 & -0.000571 & -0.000426 & -0.000227 & -0.000235 & 0.000000 & 0.000097 & 0.000131 & 0.000214 & -0.001982 \\ 
\hline
From 100 to 1,000 & 0.000758 & -0.001220 & -0.000886 & -0.000661 & -0.000356 & -0.000391 & -0.000097 & 0.000000 & 0.000049 & 0.000218 & -0.003080 \\ 
\hline
From 1,000 to 10,000 & 0.000617 & -0.001254 & -0.000911 & -0.000680 & -0.000368 & -0.000411 & -0.000131 & -0.000049 & 0.000000 & 0.000187 & -0.003170 \\ 
\hline
From 10,000 to 100,000 & -0.000144 & -0.000933 & -0.000678 & -0.000507 & -0.000280 & -0.000338 & -0.000214 & -0.000218 & -0.000187 & 0.000000 & -0.002368 \\ 
\hline
From 100,000 to infinity & 0.010290 & -0.000056 & -0.000039 & -0.000016 & 0.000079 & 0.000522 & 0.001982 & 0.003080 & 0.003170 & 0.002368 & 0.000000 \\ 
\hline
\end{tabular}
}
\caption{BIP 42: Interaction rates matrices}
\end{center}
\end{table}

\begin{table}[hbt!]
\scriptsize
\begin{center}
\rotatebox{90}{
\begin{tabular}{ |c|c|c|c|c|c|c|c|c|c|c|c| }
\hline
 & \rotatebox{270}{Control Mechanism} & \rotatebox{270}{From 0 to 0.001} & \rotatebox{270}{From 0.001 to 0.01} & \rotatebox{270}{From 0.01 to 0.1} & \rotatebox{270}{From 0.1 to 1} & \rotatebox{270}{From 1 to 10} & \rotatebox{270}{From 10 to 100} & \rotatebox{270}{From 100 to 1,000} & \rotatebox{270}{From 1,000 to 10,000} & \rotatebox{270}{From 10,000 to 100,000} & \rotatebox{270}{From 100,000 to infinity} \\ 
\hline
\multicolumn{12}{|c|}{\textbf{BIP 50 - Before}} \\
\hline
Control Mechanism & 0.000000 & -0.007518 & -0.004942 & -0.001617 & -0.003419 & -0.001500 & 0.000526 & -0.000980 & -0.001471 & -0.001771 & -0.000442 \\ 
\hline
From 0 to 0.001 & 0.007518 & 0.000000 & 0.000000 & 0.000002 & 0.000013 & 0.000124 & 0.000992 & 0.001378 & 0.001518 & 0.001007 & 0.000064 \\ 
\hline
From 0.001 to 0.01 & 0.004942 & -0.000000 & 0.000000 & 0.000002 & 0.000008 & 0.000081 & 0.000653 & 0.000900 & 0.000990 & 0.000656 & 0.000042 \\ 
\hline
From 0.01 to 0.1 & 0.001617 & -0.000002 & -0.000002 & 0.000000 & 0.000002 & 0.000026 & 0.000214 & 0.000294 & 0.000323 & 0.000214 & 0.000014 \\ 
\hline
From 0.1 to 1 & 0.003419 & -0.000013 & -0.000008 & -0.000002 & 0.000000 & 0.000054 & 0.000452 & 0.000621 & 0.000682 & 0.000451 & 0.000028 \\ 
\hline
From 1 to 10 & 0.001500 & -0.000124 & -0.000081 & -0.000026 & -0.000054 & 0.000000 & 0.000207 & 0.000257 & 0.000276 & 0.000170 & 0.000005 \\ 
\hline
From 10 to 100 & -0.000526 & -0.000992 & -0.000653 & -0.000214 & -0.000452 & -0.000207 & 0.000000 & -0.000225 & -0.000299 & -0.000303 & -0.000063 \\ 
\hline
From 100 to 1,000 & 0.000980 & -0.001378 & -0.000900 & -0.000294 & -0.000621 & -0.000257 & 0.000225 & 0.000000 & -0.000071 & -0.000192 & -0.000072 \\ 
\hline
From 1,000 to 10,000 & 0.001471 & -0.001518 & -0.000990 & -0.000323 & -0.000682 & -0.000276 & 0.000299 & 0.000071 & 0.000000 & -0.000159 & -0.000076 \\ 
\hline
From 10,000 to 100,000 & 0.001771 & -0.001007 & -0.000656 & -0.000214 & -0.000451 & -0.000170 & 0.000303 & 0.000192 & 0.000159 & 0.000000 & -0.000044 \\ 
\hline
From 100,000 to infinity & 0.000442 & -0.000064 & -0.000042 & -0.000014 & -0.000028 & -0.000005 & 0.000063 & 0.000072 & 0.000076 & 0.000044 & 0.000000 \\ 
\hline
\multicolumn{12}{|c|}{\textbf{BIP 50 - After}} \\
\hline
Control Mechanism & 0.000000 & -0.006611 & -0.003559 & -0.004262 & -0.005677 & -0.003416 & -0.000641 & -0.000407 & -0.001199 & -0.000692 & -0.000006 \\ 
\hline
From 0 to 0.001 & 0.006611 & 0.000000 & 0.000000 & 0.000002 & 0.000014 & 0.000123 & 0.000856 & 0.001391 & 0.001579 & 0.001097 & 0.000062 \\ 
\hline
From 0.001 to 0.01 & 0.003559 & -0.000000 & 0.000000 & 0.000001 & 0.000007 & 0.000064 & 0.000458 & 0.000746 & 0.000843 & 0.000587 & 0.000033 \\ 
\hline
From 0.01 to 0.1 & 0.004262 & -0.000002 & -0.000001 & 0.000000 & 0.000006 & 0.000076 & 0.000549 & 0.000893 & 0.001009 & 0.000703 & 0.000040 \\ 
\hline
From 0.1 to 1 & 0.005677 & -0.000014 & -0.000007 & -0.000006 & 0.000000 & 0.000096 & 0.000730 & 0.001188 & 0.001342 & 0.000935 & 0.000053 \\ 
\hline
From 1 to 10 & 0.003416 & -0.000123 & -0.000064 & -0.000076 & -0.000096 & 0.000000 & 0.000428 & 0.000707 & 0.000786 & 0.000550 & 0.000032 \\ 
\hline
From 10 to 100 & 0.000641 & -0.000856 & -0.000458 & -0.000549 & -0.000730 & -0.000428 & 0.000000 & 0.000081 & -0.000004 & 0.000016 & 0.000005 \\ 
\hline
From 100 to 1,000 & 0.000407 & -0.001391 & -0.000746 & -0.000893 & -0.001188 & -0.000707 & -0.000081 & 0.000000 & -0.000156 & -0.000078 & 0.000002 \\ 
\hline
From 1,000 to 10,000 & 0.001199 & -0.001579 & -0.000843 & -0.001009 & -0.001342 & -0.000786 & 0.000004 & 0.000156 & 0.000000 & 0.000035 & 0.000010 \\ 
\hline
From 10,000 to 100,000 & 0.000692 & -0.001097 & -0.000587 & -0.000703 & -0.000935 & -0.000550 & -0.000016 & 0.000078 & -0.000035 & 0.000000 & 0.000005 \\ 
\hline
From 100,000 to infinity & 0.000006 & -0.000062 & -0.000033 & -0.000040 & -0.000053 & -0.000032 & -0.000005 & -0.000002 & -0.000010 & -0.000005 & 0.000000 \\ 
\hline
\end{tabular}
}
\caption{BIP 50: Interaction rates matrices}
\end{center}
\end{table}

\begin{table}[hbt!]
\scriptsize
\begin{center}
\rotatebox{90}{
\begin{tabular}{ |c|c|c|c|c|c|c|c|c|c|c|c| }
\hline
 & \rotatebox{270}{Control Mechanism} & \rotatebox{270}{From 0 to 0.001} & \rotatebox{270}{From 0.001 to 0.01} & \rotatebox{270}{From 0.01 to 0.1} & \rotatebox{270}{From 0.1 to 1} & \rotatebox{270}{From 1 to 10} & \rotatebox{270}{From 10 to 100} & \rotatebox{270}{From 100 to 1,000} & \rotatebox{270}{From 1,000 to 10,000} & \rotatebox{270}{From 10,000 to 100,000} & \rotatebox{270}{From 100,000 to infinity} \\ 
\hline
\multicolumn{12}{|c|}{\textbf{BIP 141 - Before}} \\
\hline
Control Mechanism & 0.000000 & -0.002514 & -0.002129 & -0.002576 & -0.004140 & -0.002844 & -0.001801 & -0.000349 & -0.000502 & -0.000966 & 0.004803 \\ 
\hline
From 0 to 0.001 & 0.002514 & 0.000000 & 0.000001 & 0.000008 & 0.000040 & 0.000211 & 0.000679 & 0.001023 & 0.000973 & 0.000735 & 0.000175 \\ 
\hline
From 0.001 to 0.01 & 0.002129 & -0.000001 & 0.000000 & 0.000006 & 0.000032 & 0.000177 & 0.000574 & 0.000867 & 0.000824 & 0.000622 & 0.000150 \\ 
\hline
From 0.01 to 0.1 & 0.002576 & -0.000008 & -0.000006 & 0.000000 & 0.000028 & 0.000207 & 0.000689 & 0.001047 & 0.000995 & 0.000750 & 0.000194 \\ 
\hline
From 0.1 to 1 & 0.004140 & -0.000040 & -0.000032 & -0.000028 & 0.000000 & 0.000301 & 0.001088 & 0.001679 & 0.001593 & 0.001195 & 0.000364 \\ 
\hline
From 1 to 10 & 0.002844 & -0.000211 & -0.000177 & -0.000207 & -0.000301 & 0.000000 & 0.000617 & 0.001129 & 0.001058 & 0.000751 & 0.000600 \\ 
\hline
From 10 to 100 & 0.001801 & -0.000679 & -0.000574 & -0.000689 & -0.001088 & -0.000617 & 0.000000 & 0.000639 & 0.000562 & 0.000266 & 0.001421 \\ 
\hline
From 100 to 1,000 & 0.000349 & -0.001023 & -0.000867 & -0.001047 & -0.001679 & -0.001129 & -0.000639 & 0.000000 & -0.000069 & -0.000291 & 0.001980 \\ 
\hline
From 1,000 to 10,000 & 0.000502 & -0.000973 & -0.000824 & -0.000995 & -0.001593 & -0.001058 & -0.000562 & 0.000069 & 0.000000 & -0.000227 & 0.001893 \\ 
\hline
From 10,000 to 100,000 & 0.000966 & -0.000735 & -0.000622 & -0.000750 & -0.001195 & -0.000751 & -0.000266 & 0.000291 & 0.000227 & 0.000000 & 0.001471 \\ 
\hline
From 100,000 to infinity & -0.004803 & -0.000175 & -0.000150 & -0.000194 & -0.000364 & -0.000600 & -0.001421 & -0.001980 & -0.001893 & -0.001471 & 0.000000 \\ 
\hline
\multicolumn{12}{|c|}{\textbf{BIP 141 - After}} \\
\hline
Control Mechanism & 0.000000 & -0.004224 & -0.003724 & -0.002176 & -0.000713 & 0.000053 & -0.000204 & -0.000795 & -0.001314 & -0.001391 & 0.002153 \\ 
\hline
From 0 to 0.001 & 0.004224 & 0.000000 & 0.000002 & 0.000020 & 0.000127 & 0.000557 & 0.001546 & 0.001936 & 0.001889 & 0.001528 & 0.000184 \\ 
\hline
From 0.001 to 0.01 & 0.003724 & -0.000002 & 0.000000 & 0.000017 & 0.000112 & 0.000492 & 0.001363 & 0.001707 & 0.001665 & 0.001347 & 0.000163 \\ 
\hline
From 0.01 to 0.1 & 0.002176 & -0.000020 & -0.000017 & 0.000000 & 0.000062 & 0.000287 & 0.000796 & 0.000994 & 0.000967 & 0.000781 & 0.000105 \\ 
\hline
From 0.1 to 1 & 0.000713 & -0.000127 & -0.000112 & -0.000062 & 0.000000 & 0.000096 & 0.000255 & 0.000303 & 0.000279 & 0.000216 & 0.000096 \\ 
\hline
From 1 to 10 & -0.000053 & -0.000557 & -0.000492 & -0.000287 & -0.000096 & 0.000000 & -0.000046 & -0.000129 & -0.000196 & -0.000202 & 0.000282 \\ 
\hline
From 10 to 100 & 0.000204 & -0.001546 & -0.001363 & -0.000796 & -0.000255 & 0.000046 & 0.000000 & -0.000197 & -0.000389 & -0.000434 & 0.000796 \\ 
\hline
From 100 to 1,000 & 0.000795 & -0.001936 & -0.001707 & -0.000994 & -0.000303 & 0.000129 & 0.000197 & 0.000000 & -0.000246 & -0.000350 & 0.001021 \\ 
\hline
From 1,000 to 10,000 & 0.001314 & -0.001889 & -0.001665 & -0.000967 & -0.000279 & 0.000196 & 0.000389 & 0.000246 & 0.000000 & -0.000147 & 0.001019 \\ 
\hline
From 10,000 to 100,000 & 0.001391 & -0.001528 & -0.001347 & -0.000781 & -0.000216 & 0.000202 & 0.000434 & 0.000350 & 0.000147 & 0.000000 & 0.000839 \\ 
\hline
From 100,000 to infinity & -0.002153 & -0.000184 & -0.000163 & -0.000105 & -0.000096 & -0.000282 & -0.000796 & -0.001021 & -0.001019 & -0.000839 & 0.000000 \\ 
\hline
\end{tabular}
}
\caption{BIP 141: Interaction rates matrices}
\end{center}
\end{table}

\begin{table}[hbt!]
\scriptsize
\begin{center}
\rotatebox{90}{
\begin{tabular}{ |c|c|c|c|c|c|c|c|c|c|c|c| }
\hline
 & \rotatebox{270}{Control Mechanism} & \rotatebox{270}{From 0 to 0.001} & \rotatebox{270}{From 0.001 to 0.01} & \rotatebox{270}{From 0.01 to 0.1} & \rotatebox{270}{From 0.1 to 1} & \rotatebox{270}{From 1 to 10} & \rotatebox{270}{From 10 to 100} & \rotatebox{270}{From 100 to 1,000} & \rotatebox{270}{From 1,000 to 10,000} & \rotatebox{270}{From 10,000 to 100,000} & \rotatebox{270}{From 100,000 to infinity} \\ 
\hline
\multicolumn{12}{|c|}{\textbf{BIP 341 - Before}} \\
\hline
Control Mechanism & 0.000000 & -0.000868 & -0.000830 & -0.000288 & -0.000378 & -0.000672 & -0.000364 & -0.000464 & -0.001177 & 0.000095 & 0.000851 \\ 
\hline
From 0 to 0.001 & 0.000868 & 0.000000 & 0.000004 & 0.000030 & 0.000121 & 0.000300 & 0.000585 & 0.000688 & 0.000858 & 0.000468 & 0.000105 \\ 
\hline
From 0.001 to 0.01 & 0.000830 & -0.000004 & 0.000000 & 0.000027 & 0.000113 & 0.000283 & 0.000558 & 0.000656 & 0.000815 & 0.000449 & 0.000105 \\ 
\hline
From 0.01 to 0.1 & 0.000288 & -0.000030 & -0.000027 & 0.000000 & 0.000027 & 0.000076 & 0.000182 & 0.000212 & 0.000245 & 0.000159 & 0.000064 \\ 
\hline
From 0.1 to 1 & 0.000378 & -0.000121 & -0.000113 & -0.000027 & 0.000000 & 0.000037 & 0.000205 & 0.000236 & 0.000211 & 0.000217 & 0.000164 \\ 
\hline
From 1 to 10 & 0.000672 & -0.000300 & -0.000283 & -0.000076 & -0.000037 & 0.000000 & 0.000327 & 0.000372 & 0.000258 & 0.000395 & 0.000375 \\ 
\hline
From 10 to 100 & 0.000364 & -0.000585 & -0.000558 & -0.000182 & -0.000205 & -0.000327 & 0.000000 & -0.000025 & -0.000434 & 0.000260 & 0.000618 \\ 
\hline
From 100 to 1,000 & 0.000464 & -0.000688 & -0.000656 & -0.000212 & -0.000236 & -0.000372 & 0.000025 & 0.000000 & -0.000475 & 0.000325 & 0.000731 \\ 
\hline
From 1,000 to 10,000 & 0.001177 & -0.000858 & -0.000815 & -0.000245 & -0.000211 & -0.000258 & 0.000434 & 0.000475 & 0.000000 & 0.000729 & 0.000983 \\ 
\hline
From 10,000 to 100,000 & -0.000095 & -0.000468 & -0.000449 & -0.000159 & -0.000217 & -0.000395 & -0.000260 & -0.000325 & -0.000729 & 0.000000 & 0.000448 \\ 
\hline
From 100,000 to infinity & -0.000851 & -0.000105 & -0.000105 & -0.000064 & -0.000164 & -0.000375 & -0.000618 & -0.000731 & -0.000983 & -0.000448 & 0.000000 \\ 
\hline
\multicolumn{12}{|c|}{\textbf{BIP 341 - After}} \\
\hline
Control Mechanism & 0.000000 & 0.000107 & -0.000416 & -0.000590 & -0.000780 & -0.000489 & -0.000410 & -0.000198 & -0.000926 & -0.000479 & 0.004359 \\ 
\hline
From 0 to 0.001 & -0.000107 & 0.000000 & -0.000001 & -0.000004 & -0.000017 & -0.000042 & -0.000077 & -0.000091 & -0.000122 & -0.000059 & -0.000010 \\ 
\hline
From 0.001 to 0.01 & 0.000416 & 0.000001 & 0.000000 & 0.000012 & 0.000058 & 0.000159 & 0.000296 & 0.000352 & 0.000465 & 0.000224 & 0.000079 \\ 
\hline
From 0.01 to 0.1 & 0.000590 & 0.000004 & -0.000012 & 0.000000 & 0.000061 & 0.000211 & 0.000407 & 0.000494 & 0.000633 & 0.000304 & 0.000235 \\ 
\hline
From 0.1 to 1 & 0.000780 & 0.000017 & -0.000058 & -0.000061 & 0.000000 & 0.000229 & 0.000496 & 0.000632 & 0.000741 & 0.000353 & 0.000758 \\ 
\hline
From 1 to 10 & 0.000489 & 0.000042 & -0.000159 & -0.000211 & -0.000229 & 0.000000 & 0.000191 & 0.000339 & 0.000193 & 0.000081 & 0.001755 \\ 
\hline
From 10 to 100 & 0.000410 & 0.000077 & -0.000296 & -0.000407 & -0.000496 & -0.000191 & 0.000000 & 0.000207 & -0.000200 & -0.000119 & 0.003173 \\ 
\hline
From 100 to 1,000 & 0.000198 & 0.000091 & -0.000352 & -0.000494 & -0.000632 & -0.000339 & -0.000207 & 0.000000 & -0.000563 & -0.000299 & 0.003726 \\ 
\hline
From 1,000 to 10,000 & 0.000926 & 0.000122 & -0.000465 & -0.000633 & -0.000741 & -0.000193 & 0.000200 & 0.000563 & 0.000000 & -0.000036 & 0.005043 \\ 
\hline
From 10,000 to 100,000 & 0.000479 & 0.000059 & -0.000224 & -0.000304 & -0.000353 & -0.000081 & 0.000119 & 0.000299 & 0.000036 & 0.000000 & 0.002438 \\ 
\hline
From 100,000 to infinity & -0.004359 & 0.000010 & -0.000079 & -0.000235 & -0.000758 & -0.001755 & -0.003173 & -0.003726 & -0.005043 & -0.002438 & 0.000000 \\ 
\hline
\end{tabular}
}
\caption{BIP 341: Interaction rates matrices}
\end{center}
\end{table}

\newpage
\section{BIP Simulations: Daily Changes}
\label{Appendix:MajorEconomicBIPsSimulationsDailyChanges}
\begin{table}[hbt!]
\footnotesize
\begin{center}
\begin{tabular}{ |c|c|c|c|c| }
\hline
 & \rotatebox{90}{Mean} & \rotatebox{90}{Standard Deviation } & \rotatebox{90}{Lower Bound} & \rotatebox{90}{Upper Bound} \\
\hline
Total & 0.079836 & 0.053010 & 0.026826 & 0.132846 \\
\hline
Control Mechanism & 0.016296 & 0.010104 & 0.006192 & 0.026400 \\
\hline
From 0 to 0.001 & 0.004105 & 0.003613 & 0.000492 & 0.007718 \\
\hline
From 0.001 to 0.01 & 0.003537 & 0.004168 & -0.000631 & 0.007705 \\
\hline
From 0.01 to 0.1 & 0.003883 & 0.003561 & 0.000322 & 0.007444 \\
\hline
From 0.1 to 1 & 0.003436 & 0.003161 & 0.000275 & 0.006597 \\
\hline
From 1 to 10 & 0.003392 & 0.002262 & 0.001130 & 0.005654 \\
\hline
From 10 to 100 & 0.005362 & 0.003902 & 0.001460 & 0.009264 \\
\hline
From 100 to 1,000 & 0.007004 & 0.004934 & 0.002070 & 0.011938 \\
\hline
From 1,000 to 10,000 & 0.007929 & 0.004911 & 0.003018 & 0.012840 \\
\hline
From 10,000 to 100,000 & 0.007163 & 0.004708 & 0.002455 & 0.011871 \\
\hline
From 100,000 to infinity & 0.017724 & 0.022793 & -0.005069 & 0.040517 \\
\hline
\end{tabular}
\caption{Benchmark statistics for the interaction rate differences computed based on the daily simulations in the period between 11-02-2012 and 19-01-2020}
\end{center}
\end{table}

\end{document}